\documentclass[leqno,12pt]{report}
\usepackage{amsmath}
\usepackage[psamsfonts]{amssymb}
\usepackage{amsthm}
\usepackage[mathscr,psamsfonts]{eucal}
\usepackage{cmmib57}
\usepackage{diagrams}
\input{ueur.fd}\input{ueur57.fd}
\DeclareMathAlphabet{\eurm}{U}{eur}{m}{n}
\DeclareMathAlphabet{\eubf}{U}{eur}{b}{n}
\DeclareFontFamily{U}{UWCyr}{}
\DeclareFontShape{U}{UWCyr}{m}{n}{%
  <5> <6> <7> <8> <9>
  <10> <10.95> <12> <14.4> <17.28> <20.74> <24.88> wncyr10
  }{}
\DeclareFontShape{U}{UWCyr}{m}{it}{%
  <5> <6> <7> <8> <9>
  <10> <10.95> <12> <14.4> <17.28> <20.74> <24.88> wncyi10
  }{}
\DeclareFontShape{U}{UWCyr}{m}{sc}{%
  <5> <6> <7> <8> <9>
  <10> <10.95> <12> <14.4> <17.28> <20.74> <24.88> wncysc10
  }{}
\DeclareFontShape{U}{UWCyr}{b}{n}{%
  <5> <6> <7> <8> <9>
  <10> <10.95> <12> <14.4> <17.28> <20.74> <24.88> wncyb10
  }{}
\DeclareMathAlphabet{\cyrm}{U}{UWCyr}{m}{n}
\DeclareMathAlphabet{\cyit}{U}{UWCyr}{m}{it}
\DeclareMathAlphabet{\cysc}{U}{UWCyr}{m}{sc}
\DeclareMathAlphabet{\cybf}{U}{UWCyr}{b}{n}

\newcommand{\mysec}[1]{\section{#1}}
\newcommand{\mysecn}[1]{\section*{#1}}

\newcounter{assump}
\newtheorem{Assumption}{\indent Assumption}[assump]
\newtheorem{Definition}{\indent Definition}[section]
\newtheorem{Lemma}{\indent Lemma}[section]
\newtheorem{Proposition}{\indent Proposition}[section]
\newtheorem{Theorem}{\indent Theorem}[section]
\newtheorem{Corollary}{\indent Corollary}[section]
\newtheorem{Remark}{\indent Remark}[section]
\newtheorem{Example}{\indent Example}[section]
\newcommand{\bAs}{\begin{Assumption}\em}
\newcommand{\eAs}{\end{Assumption}}
\newcommand{\bDf}{\begin{Definition}\em}
\newcommand{\eDf}{\end{Definition}}
\newcommand{\bLm}{\begin{Lemma}}
\newcommand{\eLm}{\end{Lemma}}
\newcommand{\bPr}{\begin{Proposition}}
\newcommand{\ePr}{\end{Proposition}}
\newcommand{\bTh}{\begin{Theorem}}
\newcommand{\eTh}{\end{Theorem}}
\newcommand{\bCr}{\begin{Corollary}}
\newcommand{\eCr}{\end{Corollary}}
\newcommand{\bRm}{\begin{Remark}\em}
\newcommand{\eRm}{\end{Remark}}
\newcommand{\bEx}{\begin{Example}\em}
\newcommand{\eEx}{\end{Example}}
\newcommand{\bPf}{\begin{proof}[\noindent\indent{\sc Proof}]}
\newcommand{\ePf}{\renewcommand{\qedsymbol}{}\end{proof}}
\newcommand{\bEq}{\begin{eqnarray}}
\newcommand{\eEq}{\end{eqnarray}}
\newcommand{\beq}{\begin{eqnarray*}}
\newcommand{\eeq}{\end{eqnarray*}}
\newcommand{\bdg}{\beq\begin{diagram}}
\newcommand{\edg}{\end{diagram}\eeq}
\newcommand{\bDg}{\bEq\begin{diagram}}
\newcommand{\eDg}{\end{diagram}\eEq}
\newcommand{\ben}{\begin{enumerate}}
\newcommand{\een}{\end{enumerate}}
\newcommand{\btb}{\begin{tabbing}}
\newcommand{\etb}{\end{tabbing}}
\def\QED{\hskip0.1em\hfill\null\ \null\nobreak\hfill\kern3pt\vbox
{\hrule\hbox
   {\vrule\kern1pt\vbox{\kern1.7pt\hbox{$\scriptscriptstyle{QED}$}
    \kern0.2pt}\kern1pt\vrule}\hrule}}

\def\END{\hskip0.1em\hfill\null\ \null\nobreak\hfill\kern3pt\vbox
{\hrule\hbox
   {\vrule\kern1pt\vbox{\kern1.7pt\hbox{$\,\,\,\vspace{5pt}$}
    \kern0.2pt}\kern1pt\vrule}\hrule}}
\newcommand{\ie}{i.e$.$ }
\newcommand{\R}{I\!\!R}

\newcommand{\mto}{\mapsto}

\newcommand{\der}{\partial}

\DeclareMathOperator{\im}{im}

\DeclareMathOperator{\id}{id}
\DeclareMathOperator{\byd}{\,{\raisebox{.1ex}{$\eurm :$}{\eurm =}}\,}

\newcommand{\drs}{\oplus}
\newcommand{\sub}{\subset}
\newcommand{\sym}{\odot}
\newcommand{\wed}{\wedge}

\newcommand{\com}{\!\circ\!}
\newcommand{\con}{\,\lrcorner\,}
\newcommand{\ten}{\!\otimes\!}

\newcommand{\ucar}[1]{\underset{#1}{\times}}

\newcommand{\owed}[1]{\overset{#1}{\wedge}}
\newcommand{\osym}[1]{\overset{#1}{\odot}}

\newcommand{\oset}[2]{\overset{#1}{#2}}

\newcommand{\olin}[1]{\overline{#1}}
\newcommand{\ulin}[1]{\underline{#1}}
\newcommand{\alp}{\alpha}
\newcommand{\bet}{\beta}
\newcommand{\gam}{\gamma}

\newcommand{\eps}{\epsilon}

\newcommand{\lam}{\lambda}

\newcommand{\ome}{\omega}
\newcommand{\Gam}{\Gamma}

\newcommand{\vartht}{\vartheta}

\newcommand{\bI}{\boldsymbol{I}}

\newcommand{\bU}{\boldsymbol{U}}
\newcommand{\bV}{\boldsymbol{V}}

\newcommand{\bX}{\boldsymbol{X}}
\newcommand{\bY}{\boldsymbol{Y}}

\newcommand{\cA}{\mathcal{A}}
\newcommand{\cB}{\mathcal{B}}
\newcommand{\cC}{\mathcal{C}}

\newcommand{\cE}{\mathcal{E}}

\newcommand{\cP}{\mathcal{P}}

\newcommand{\cS}{\mathcal{S}}

\newcommand{\BN}{{\mathbb{N}}}

\newcommand{\introductionname}{Introduction}
\newcommand{\myintro}{
  \section*{\introductionname}
  \markboth{\introductionname}{\introductionname}
  \addcontentsline{toc}{section}{\hspace*{1.5em}\introductionname}
}
\newcommand{\up}{\underline{p}}
\newcommand{\uq}{\underline{q}}
\newcommand{\us}{\underline{s}}
\newcommand{\ut}{\underline{t}}
\newcommand{\D}{\cyrm{d}}

\newcommand{\myskip}{\vspace*{8pt}}
\let\mytext=\text
\newcommand{\For}[1]{\overset{#1}{\Lambda}}
\newcommand{\Con}[1]{\overset{#1}{\cal{C}}}
\newcommand{\Hor}[1]{\overset{#1}{\cal{H}}}
\newcommand{\Var}[1]{\overset{#1}{\cal{V}}}
\newcommand{\Thd}[1]{\overset{#1}{\Theta}}
\title{{\bf On finite order variational sequences}}
\author{Raffaele Vitolo\thanks{
This paper has been partially supported by INdAM \lq
F. Severi\rq through a senior research fellowship, GNFM of CNR, MURST,
Universities of Florence and Lecce.}
\\{\small Department of Mathematics ``E. De Giorgi", 
University of Lecce}
\\{\small Via per Arnesano, 73100 Lecce, Italy}
\\{\small E--mail: vitolo@ilenic.unile.it}
\\{\small Web: www.dm.unile.it/~mongelli/docenti/}}
\date{}
\begin{document}

\maketitle

\begin{abstract}

We discuss intrinsic aspects of Krupka's approach to finite--order variational 
sequences. We give intrinsic isomorphisms of the quotient subsheaves of the 
short finite--order variational sequence with sheaves of forms on jet spaces 
of suitable order, obtaining a new finite--order (short exact) variational 
sequence which is made by sheaves of polynomial differential operators.  
Moreover, we present an intrinsic formulation for the Helmholtz condition of 
local variationality using a technique introduced by Kol\' a\v r that we have 
adapted to our context. Finally, we provide the minimal order solution to the
inverse problem of the calculus of variations and a solution of the problem of
the variationally trivial Lagrangian.

\myskip

\noindent {\bf Key words}: Fibred manifold, jet space, variational sequence, 
Euler--Lagrange morphism, Helmholtz morphism.

\noindent {\bf 1991 MSC}: 58A12, 58A20, 58E30, 58G05.

\end{abstract}

\newpage

\tableofcontents

\myintro

It is known that there exist several geometric formulations of the
variational calculus. They are inspired by a geometrical version of
the Hamilton's principle of least action, stated on a fibred manifold.
See, for example, \cite{Gar74,GoSt73,Kru73,Tul75}, and for further
developments \cite{Cos94,Cra81,Fer83,FeFr82,GaMu82,Kol83,Kru83,MaMo83b,Sau89}.
In these papers the leading idea is that one can introduce the
variational calculus in a purely differential context. See the
Appendix for an introduction to this formalism.

Variational sequences go a step forward according to this 
guideline \cite{AnDu80,Kup80,OlSh78,Tak79,Tul77,Tul80,Vin77,Vin78}. The basic
idea is to interpret the passages from a Lagrangian to its 
Euler--Lagrange morphism and from an Euler--Lagrange morphism to its 
conditions of local variationality (Helmholtz' conditions)
as morphisms of an exact sequence, namely the variational sequence.
This is the framework where a lot of problems and ambiguities of geometrical 
formulations of Lagrangian field theories and mechanics can be solved. 
See \cite{Tra96} for a discussion of these problems.

\myskip

But in \cite{Kup80,OlSh78,Tak79,Tul77,Tul80,Vin77,Vin78} the 
variational sequence is built over the space of infinite jets of a fibred 
manifold. This procedure is suggested by the relatively simple structure of 
such spaces. Only in \cite{AnDu80} there is a partial construction on 
finite order jets.

This paper deals with Krupka's setting of variational sequence on 
finite--order jet spaces \cite{Kru90} (for further developments, see
\cite{Kru93,Kru95a,Kru95b,KrMu99}). 
The finite--order variational sequence is produced when one quotients the de
Rham sequence on a finite--order jet space by means of an intrinsically defined
subsequence. The choice of this subsequence is inspired by the variational
calculus; it is made by forms which do not contribute to action--like
integrals. 

Several papers investigated problems arising from the above construction 
\cite{Gri99a,Gri99b,Kas99,Mus95,MuKr99,Ste95}. But all of them are not
concerned with the intrinsic aspects of the problems that they face.

\myskip

In this paper (and in
\cite{FrPaVi99,FrPaVi99b,Vit95,Vit96a,Vit97,Vit98,Vit96c,Vit99b})
our leading idea is to analyse Krupka's variational
sequence by means of intrinsic techniques on jet spaces. Namely, we will
use the structure form on jet spaces \cite{MaMo83a} and the geometric
version of the first variation formula by \cite{Kol83}.

\myskip

In \cite{Vit95,Vit96a}, we analysed the particular case of the 
first--order variational sequence on a fibred manifold whose base is 
$1$--dimensional.  This was done in order to reduce technical difficulties.  
Here, we analyse the most general situation, \ie the $r$--th order 
variational sequence based on a fibred manifold, without any restriction on 
the dimension of the base.  We give isomorphisms of the quotient sheaves of 
the variational sequence with subsheaves of the sheaves of forms on a jet space
of suitable order. This order is always found as the minimal among all
possible  candidates; this aspect is not present in the infinite jet formalism.

We give a characterisation of the local conditions of local variationality.  
More precisely, it is known \cite{Bau82,Kru90} that there exists a locally 
defined geometric object, namely the Helmholtz morphism, whose vanishing is 
equivalent to the local conditions of local variationality 
\cite{And86,GiMa90,LaTu77,Kru90,Ton69}.  We show that the Helmholtz morphism 
is intrinsically characterised by means of the Euler--Lagrange morphism. This
issue is also present in \cite{Gri99a}, with a slightly different proof. In
this way, we obtain that the variationality conditions are global and
intrinsic. This fact is also due to the intrinsic nature of the variational
sequence. Moreover, we obtain an intrinsic geometrical object which plays a
role analogous to the role of the momentum of a Lagrangian.

Finally, we obtain a finite--order (short and exact) variational sequence, 
whose sheaves are constituted by polynomial differential operators. This 
allows us to give a solution of the problem of the minimal order Lagrangian.  
Indeed, given a locally variational 
Euler--Lagrange morphism $\eps$ of order $s$, the theory of infinite order 
variational sequences yields the existence of a (local) Lagrangian of order 
$s$ inducing $\eps$.  But the finite order variational sequence provides the
minimal order Lagrangian inducing $\eps$.
The solution of this long--standing problem of the calculus of variations was 
announced (but not given) by Anderson \cite{And86,And92,AnTh92}. The finite
order variational sequence yields a proof of this condition which is of
\lq structural\rq\ nature, rather than of \lq computational\rq\ nature.
We also identify each minimal order variationally trivial
Lagrangian by a very simple intrinsic technique. Our result agrees with local
results from \cite{Gri99b,KrMu99}. 

\myskip

We notice that a short version of this report has already been published in
\cite{Vit97}. The results of this paper has been improved and completed
ever since. Indeed, it has been shown \cite{Vit98,Vit96c} that Krupka's
approach to variational sequences can be equivalently reformulated in the
context of $\cC$--spectral sequences \cite{Vin77,Vin78,Vin84}, both in the
finite and infinite order case. Also, $\cC$--spectral sequences allow to
extend the finite order formalism to jets of submanifolds and
differential equations, and Green--Vinogradov formula \cite{Vin84} allows us to
represent each quotient space of the variational sequence in an
intrinsic way \cite{Vit99b}. Finally, symmetries has been fitted into Krupka's
framework \cite{FrPaVi99,FrPaVi99b}, recovering old results and stating some
new results.

\myskip

We hope that our work could serve as a tool to both mathematical and
theoretical physicists for a deeper understanding of Lagrangian formalism.

\mysecn{Preliminaries}

In this paper, manifolds and maps between manifolds are $C^{\infty}$. All
morphisms of fibred manifolds (and hence bundles) will be morphisms over 
the identity of the base manifold, unless otherwise specified.

\myskip

Let $V$ be a vector space such that $\dim V = n$. Suppose that
$V=W_1\drs W_2$, with $p_1:V\to W_1$ and $p_2:V\to W_2$ the related
projections. Then, we have the splitting
\bEq
 \label{wed split}
   \owed{m}V=\bigoplus_{k+h=m}\owed{k}W_1\wed\owed{h}W_2~,
\eEq
where $\owed{k}W_1\wed\owed{h}W_2$ is the subspace of $\owed{m}V$ generated by
the wedge products of elements of $\owed{k}W_1$ and $\owed{h}W_2$.

There exists a natural inclusion $\osym{k}\,L(V,V) \sub L(\owed{k}V,
\owed{k}V)$. Then, the projections $p_{k,h}$ related to the
above splitting turn out to be the maps
\beq
   p_{k,h}=\binom{k}{p}\osym{k}p_1\osym{}\osym{h}p_2
   :\owed{m}V\to\owed{k}W_1\wed\owed{h}W_2~.
\eeq

Let $V' \sub V$ be a vector subspace, and set $W'_{1} \byd p_{1}(V')$,
$W'_{2} \byd p_{2}(V')$.  Then we have 
\bEq\label{wed split and ssp}
V' \sub W'_{1} \drs W'_{2} \,,
\eEq
but the inclusion, in general, is not an equality.

\myskip

As for sheaves, we will use the definitions and the main results given in
\cite{Wel80}. In particular, we will be concerned only with sheaves
of $\R$--vector spaces. Thus, by \lq sheaf morphism\rq\  we will mean morphism
of sheaves of $\R$--vector spaces. 

Let $\cP$ be a presheaf over a topological space $X$. We will denote
by $\olin{\cP}$ the sheaf generated by $\cP$ in the sense of 
\cite{Wel80}. This means that $\olin{\cP}$ is a completion of $\cP$ 
with respect to the gluing axiom. We will denote by
$\cP_U$ the set of sections of $\cP$ defined on the open
subset $U\subset X$. The sum between two local sections $\alp\in\cP$ and
$\bet\in\cP$ will be defined on the intersection of their domain of
definition. If $\cA$, $\cB$ are two subpresheaves of a presheaf $\cP$, then
the wedge product $\cA\wed\cB$ is defined to be the subpresheaf of sections of 
$\owed{2}\cP$ generated by wedge products of sections of $\cA$ and $\cB$.

Let $\cS$ be a sheaf. We recall that $\cS$ is said to be {\em soft\/}
if each section defined on a closed subset $C\subset X$ can be extended
to a section defined on any open subset $U$ such that $C\subset U$. 
Moreover, $\cS$ is said to be {\em fine\/} if it admits a partition of unity.
A fine sheaf is also a soft sheaf. We recall also that a sequence of 
sheaves over $X$ is said to be exact if it is locally exact 
(see \cite{Wel80} for a more precise definition).
Finally, we recall that the sheaf of sections of a vector bundle is a 
fine sheaf, hence a soft sheaf.

\myskip

{\bf Acknowledgements.} 
I would like to thank I. Kol\'a\v r, D. Krupka, M. Modugno, and J. \v
Stef\'anek for helpful suggestions.

The commutative diagrams are produced by Paul Taylor's \texttt{diagrams}
macro package, available in CTAN in
\texttt{TeX/macros/generic/diagrams/taylor}.
\chapter{Jet spaces}

In this chapter we recall some facts on jet spaces. We start with the 
definition of jet space, then we introduce the contact maps. We study the 
natural sheaves of forms on jet spaces which arise from the fibring and the
contact maps. Finally, we introduce the horizontal and vertical 
differential of forms on jet spaces.

\section{Jet spaces}

Our framework is a fibred manifold
\beq
\pi : \bY \to \bX \,,
\eeq
with $\dim \bX = n$ and $\dim \bY = n+m$.

We deal with the tangent bundle $T\bY \to \bY$, the tangent
prolongation $T\pi : T\bY \to T\bX$ and the vertical bundle $V\bY \to \bY$.

Moreover, for $0 \leq r$, we are concerned with the $r$--jet space $J_r\bY$;
in particular, we set $J_0\bY \equiv \bY$. We recall the natural fibrings
\beq
\pi^r_s : J_r\bY \to J_s\bY \,, \qquad \pi^r : J_r\bY \to \bX \,, 
\eeq
and the affine bundle
\begin{align*}
&\pi^r_{r-1} : J_r\bY \to J_{r-1}\bY
\\
\intertext{associated with the vector bundle}
&\hphantom{t^r_{r-1} :} 
\sym^r T^*\bX \underset{J_{r-1}\bY}{\otimes} V\bY \to J_{r-1}\bY \,, 
\end{align*}
for $0 \leq s \leq r$. A detailed account of the theory of jets can be found 
in \cite{MaMo83a,Kup80,Sau89}.

\myskip

Charts on $\bY$ adapted to the fibring are denoted by $(x^\lam ,y^i)$.  Greek 
indices $\lam ,\mu ,\dots$ run from $1$ to $n$ and label base coordinates, 
Latin indices $i,j,\dots$ run from $1$ to $m$ and label fibre coordinates, 
unless otherwise specified.  We denote by $(\der_\lam ,\der_i)$ and $(d^\lam 
,d^i)$, respectively, the local bases of vector fields and $1$--forms on $\bY$ 
induced by an adapted chart. 

We denote multi--indices of dimension $n$ by underlined latin letters such as
$\up = (p_1, \dots, p_n)$, with $0 \leq p_1, \dots, p_n$; by identifying the
index $\lam $ with a multi--index according to
\beq
\lam \simeq (p_1, \dots, p_{\lam}, \dots, p_n) \equiv (0, \dots, 1, \dots, 0) \,,
\eeq
we can write
\beq
\up + \lam = (p_1, \dots, p_{\lam} + 1, \dots, p_n) \,.
\eeq
We also set
$|\up | \byd p_{1} + \dots + p_{n}$ and $\up ! \byd p_{1}! \dots p_{n}!$.

The charts induced on $J_r\bY$ are denoted by $(x^0,y^i_{\up})$, with $0 \leq 
|\up| \leq r$; in particular, if $|\up| = 0$, then we set $y^i_{\ulin{0}} 
\equiv y^i$.  The local vector fields and forms of $J_r\bY$ induced by the 
fibre coordinates are denoted by $(\der^{\up}_i)$ and $(d^i_{\up})$, $0 \leq 
|\up| \leq r, 1 \leq i \leq m$, respectively.

\section{Contact maps}

A fundamental role is played in the theory of variational sequences by the 
``contact maps" on jet spaces (see \cite{MaMo83a}).  Namely, for $1 \leq r$, 
we consider the natural injective fibred morphism over $J_r\bY \to J_{r-1}\bY$
\beq
\D_r : J_r\bY \ucar{\bX} T\bX \to TJ_{r-1}\bY \,,
\eeq
and the complementary surjective fibred morphism 
\beq
\vartht_r : J_r\bY \ucar{J_{r-1}\bY} TJ_{r-1}\bY \to VJ_{r-1}\bY \,,
\eeq
whose coordinate expression are 
\begin{align}\nonumber
\D_r &= d^\lam\ten {\D_r}_\lam = d^\lam\ten 
(\der_\lam + y^j_{\up+\lam}\der_j^{\up}) \,,
\qquad
0 \leq |\up| \leq r-1 ,
\\ \nonumber
\vartht_r &= \vartht^j_{\up}\ten\der_j^{\up} = 
(d^j_{\up}-y^j_{{\up}+\lam}d^\lam)
\ten\der_j^{\up} \,,
\qquad
0 \leq |\up| \leq r-1 \,.
\end{align}
We stress that
\begin{gather}
 \label{d e vart 1}
   \cyrm{d}_r\con\vartht_r=\vartht_r\con\cyrm{d}_r=0
\\
 \label{d e vart 2}
   (\vartht_r)^2=\vartht_r\qquad\qquad(\cyrm{d}_r)^2=\cyrm{d}_r
\end{gather}

The transpose of the map $\vartht_r$ is the injective fibred morphism over
$J_r\bY \to J_{r-1}\bY$
\beq
\vartht_r^* : J_r\bY \ucar{J_{r-1}\bY} V^*J_{r-1}\bY \to T^*J_r\bY \,.
\eeq
We have the remarkable vector subbundle
\bEq
 \label{contact bundles}
   \im\vartht_r^*\subset J_r\bY\ucar{J_{r-1}\bY}T^*J_{r-1}\bY
   \subset T^*J_r\bY ~,
\eEq   
and, for $0\leq t\leq s\leq r$, the fibred inclusions
\bEq
 \label{contact subbundles}
   J_r\bY\ucar{J_t\bY}\im\vartht_t^*\subset
   J_r\bY\ucar{J_s\bY}\im\vartht_s^*\subset\im\vartht_r^*~.
\eEq
The above vector subbundle $\im\vartht_r^*$ yields the splitting  
\cite{MaMo83a}
\bEq
 \label{jet connection}
  J_r\bY\ucar{J_{r-1}\bY}T^*J_{r-1}\bY =\left(
  J_r\bY\ucar{J_{r-1}\bY}T^*\bX\right) \oplus\im \vartht_r^*~.
\eEq
\section{Distinguished sheaves of forms}

We are concerned with some distinguished sheaves of forms on jet spaces.

\bRm\label{cohom. of jets} 
The manifold $\bY$ is a differentiable retract of $J_r\bY$, hence the de Rham 
cohomologies of $\bY$ and $J_r\bY$ are isomorphic.  Therefore, we reduce 
sheaves on $J_r\bY$ to sheaves on $\bY$ by considering for each sheaf $\cS$ on 
$J_r\bY$ the sheaf induced by $\cS$ by restricting to the tube topology on 
$J_r\bY$, \ie , the topology generated by open sets of the kind $\left( 
{\pi_0^r}\right)^{-1}(\bU)$, with $\bU\subset\bY$ open in $\bY$.  So, from now 
on, the sheaves of forms on $J_r\bY$ and the related subsheaves will be 
considered as sheaves over the topological space $\bY$ of the above kind.\END
\eRm

Let $0 \leq k$.

\begin{enumerate}
\item
First of all, for $0 \leq r $, we consider the standard sheaf $\For{k}_r$
of $k$--{\em forms\/} on $J_r\bY$
\beq
   \alp : J_r\bY \to \owed{k}T^*J_r\bY \,.
\eeq
\item
Then, for $0 \leq s \leq r $, we consider the sheaves $\Hor{k}_{(r,s)}$ and
$\Hor{k}_r$ of {\em horizontal forms\/}, \ie of local fibred morphisms over 
$J_r\bY \to J_s\bY$ and $J_r\bY \to \bX$ of the type
\beq
\alp : J_r\bY \to \owed{k}T^*J_s\bY
\qquad \mytext{and} \qquad
\bet : J_r\bY \to \owed{k}T^*\bX \,,
\eeq
respectively. In coordinates, if $0 < k \leq n$, then
\begin{align*}
&\alp = \alp  
{_{i_1 \dots i_{h} }^{\up_1 \dots \up_{h}}}
{_{\lam_{h+1} \dots \lam_{k}}} \,
d^{i_1}_{\up_1}\wed\dots\wed d^{i_{h}}_{\up_{h}}\wed 
d^{\lam_{h+1}} \wed\dots\wed d^{\lam_{k}}
\\
&\bet = \bet_{\lam_{1} \dots \lam_{k}} \, d^{\lam_{1}}
\wed \dots \wed d^{\lam_{k}} \,;
\end{align*}
if $k>n$, then
\beq
\alp = \alp  
{_{i_1 \dots i_{k-n+l} }^{\up_1 \dots \up_{k-n+l}}}
{_{\lam_{l+1} \dots \lam_{n}}} \,
d^{i_1}_{\up_1}\wed\dots\wed d^{i_{k-n+l}}_{\up_{k-n+l}}\wed 
d^{\lam_{l+1}} \wed\dots\wed d^{\lam_{n}} \,,
\eeq
Here, the coordinate functions are sections of $\For{0}_{r}$, and the 
indices' range is $0 \leq |\up_j| \leq s$, $0 \leq h \leq k$ and $0 \leq l \leq 
n$.  We remark that, in the coordinate expression of $\alp$, the indices 
$\lam_j$ are suppressed if $h=k$ or $l=n$, and the indices ${_{i_j}^{\up_j}}$ 
are suppressed if $h = 0$.

Clearly $\Hor{k}_{(r,r)} = \For{k}_r$ and $\Hor{k}_r = 0$ for $k>n$.

If $0\leq q\leq r$, then pull--back by $\pi^r_q$ yields
the sheaf inclusions
\begin{gather*}
   \Hor{k}_q\simeq
   {\pi^r_q}^*\Hor{k}_q\subset\Hor{k}_r\subset\Hor{k}_{(r,t)}\subset
   \Hor{k}_{(r,s)}\subset\For{k}_r~,
\\
   \For{k}_s\simeq
   {\pi^r_s}^*\For{k}_s\subset\Hor{k}_{(r,s)}\subset\For{k}_r~.
\end{gather*}

   The above inclusions are proper inclusions if $t< s< r$ and $q<r$. 
   Indeed, not all sections of the pull--back of a bundle (like
   $J_r\bY\ucar{J_s\bY}T^*J_s\bY$)
   are the pull--back of some section of the bundle itself. In fact, we 
   deal with two different operations: pull--back of bundles and pull--back of
   sections (forms).
\item
   For $0 \leq s < r$, we consider the subsheaf $\Con{k}_{(r,s)}
   \sub \Hor{k}_{(r,s)}$ of {\em contact\/} forms, \ie 
   of local fibred morphisms over $J_r\bY \to J_{s}\bY$ of the type
   \beq
      \alp : J_r\bY\to \owed{k} \im\vartht_{s+1}^* 
      \sub \owed{k}T^*J_{s}\bY \,.
   \eeq

   Due to the injectivity of $\vartht_{s+1}^*$, the subsheaf 
   $\Con{k}_{(r,s)}$ turns
   out to be the sheaf of local fibred morphisms $\alp \in \Hor{k}_{(r,s)}$
   which factorise as $\alp = \owed{k}\vartht_{s+1}^* \com \Tilde\alp$, 
   through the composition
   \bdg
      J_r\bY & 
      \rTo^{\Tilde\alp}  &
      J_{s+1}\bY \ucar{J_{s}\bY} \owed{k} V^*J_{s}\bY &
      \rTo^{\owed{k}\vartht_{s+1}} &
      \owed{k} T^*J_{s}\bY \,.
   \edg
   Thus, $\alp \in \Con{k}_{(r,s)}$ if and only if its coordinate expression 
   is of the type
   \beq
   \alp = \alp{_{i_{1}\dots i_{k}}^{\up_{1}\dots \up_{k}}} \,
   \vartht^{i_{1}}_{\up_{1}} \wed\dots\wed \vartht^{i_{k}}_{\up_{k}}
    \qquad 0 \leq |\up_{1}|,\dots ,|\up_{k}| \leq s \,,
   \eeq
   with $\alp{_{i_{1}\dots i_{k}}^{\up_{1}\dots \up_{k}}} \in \For{0}_r$.

   If $0\leq s < r \leq r'$, $s \leq s'$, then we have the inclusions (see 
   \eqref{contact bundles} and \eqref{contact subbundles})
   \beq
      \Con{k}_{(r,s)}\subset\Con{k}_{(r',s')}~.
   \eeq
\item
   Furthermore, we consider the subsheaf $\Hor{k}{_r^P} \sub
   \Hor{k}_{r}$ of local fibred morphisms $\alp \in \Hor{k}_r$ such 
   that $\alp$ is a polynomial fibred morphism over $J_{r-1}\bY \to\bX$
   of degree $k$.
   Thus, in coordinates, $\alp \in \Hor{k}{_r^P}$ if and only if
   $\alp _{\lam_{1}, \dots ,\lam_{k}} : J_r\bY \to \R$ 
   is a polynomial map of degree $k$ with respect to the 
   coordinates $y^i_{\up}$, with $|\up| = r$.
\item
   Finally, we consider the subsheaf $\Con{k}{_r} \sub
   \Con{k}_{(r+1,r)}$ of local fibred morphisms 
   $\alp \in \Con{k}_{(r+1,r)}$ such 
   that $\Tilde\alp$ projects down on $J_{r}\bY$.
   Thus, in coordinates, $\alp \in \Con{k}{_r}$ if and only if
   $\alp{_{i_{1}\dots i_{k}}^{\up_{1}\dots \up_{k}}} \in \For{0}_{r}$.
\end{enumerate}
\section{Main splitting}

The maps $\cyrm{d}_r$ and $\vartht_r$ induce two important derivations of 
degree $0$ (see \cite{Sau89,Cos94}), namely the interior products by 
$\cyrm{d}_r$ and $\vartht_r$
\beq
i_h \equiv i_{\cyrm{d}_{r+1}} : \For{k}_{r} \to \For{k}_{r+1} \,,
\qquad
i_v \equiv i_{\vartht_{r+1}} : \For{k}_{r} \to\For{k}_{r+1} \,,
\eeq
which make sense taking into account the natural 
inclusions $J_r\bY\ucar{\bX}T^*\bX\subset T^*J_r\bY$ and $VJ_r\bY\subset
TJ_r\bY$.

The fibred splitting \eqref{jet connection} yields a fundamental sheaf 
splitting.
\bLm
   We have the splitting
   \beq
      \Hor{1}_{(r+1,r)}=\Hor{1}_{r+1}\oplus\Con{1}_{(r+1,r)}~,
   \eeq
   where the projection on the first factor and on the second factor 
   are given, respectively, by
   \begin{gather*}
      H :\Hor{1}_{(r+1,r)}\to\Hor{1}_{r+1} :\alp\mto i_h\alp~,
   \\
      V :\Hor{1}_{(r+1,r)}\to\Con{1}_{(r+1,r)} :\alp\mto i_v\alp~.
   \end{gather*}
\eLm

If $\alp\in\Hor{1}_{(r+1,r)}$ has the coordinate expression
$\alp =\alp_\lam d^\lam +\alp_i^{\up}d^i_{\up}$ ($0 \leq \up \leq r$), then
\beq
   H(\alp ) =(\alp_\lam+y^i_{\up}\alp_i^{\up})\ d^\lam ~,\qquad\quad
   V(\alp ) =\alp_i^{\up}\vartht^i_{\up}~.
\eeq
\bPr
 \label{graded}
   The above splitting of $\Hor{1}_{(r+1,r)}$ induces the splitting
   \beq
      \Hor{k}_{(r+1,r)}=\bigoplus_{l=0}^k
      \Con{k-l}_{(r+1,r)}\wed\Hor{l}_{r+1}
   \eeq
   (see Preliminaries).
\ePr

We recall that, in the above splitting, direct summands with $l>n$ vanish.

We set $H$ to be the projection of the above splitting on the factor with the
highest degree of the horizontal factor. 

\bPr
If $k \leq n$, then we have
\beq
H:\Hor{k}_{(r+1,r)} \to \Hor{k}_{r+1} : 
\alp \mto \frac{1}{k!} \square^k \cyrm{d}_{r+1} (\alp) \,;
\eeq
if $k > n$, then we have
\beq
H:\Hor{k}_{(r+1,r)}\to\Con{k-n}_{(r+1,r)}\wed\Hor{n}_{r+1} :
\alp\mto \frac{1}{(k-n)! \, n!}\left(\square^{k-n}\vartht_{r+1}
\square^n \cyrm{d}_{r+1}\right)(\alp )\,.
\eeq
\ePr

\bPf
See Preliminaries.\QED
\ePf

We set also
\beq
V \byd Id - H
\eeq
to be the projection complementary to $H$.

\bRm\label{coord. expr. of h} 
If $k \leq n$, then we have the coordinate expression
\beq
H(\alp) = 
y^{i_1}_{\up_{1}+\lam_{1}} \dots y^{i_{h}}_{\up_{h}+\lam_{h}}
\alp  
{_{i_1 \dots i_{h} }^{\up_1 \dots \up_{h}}}
{_{\lam_{h+1} \dots \lam_{k}}}
d^{\lam_{1}} \wed\dots\wed d^{\lam_{k}} \,,
\eeq
with $0 \leq h \leq k$. If $k>n$, then we have
\begin{align*}
& H(\alp) = 
\sum y^{j_1}_{\uq_{1}+\lam_{1}} \dots y^{j_{l}}_{\uq_{l}+\lam_{l}}
\alp  
{_{i_1 \widehat{\dots} \, i_{k-n+l} \, j_{1} \dots j_{l}}
^{\up_1 \widehat{\dots} \up_{k-n+l} \uq_{1} \dots \uq_{l}}}
{_{\lam_{l+1} \dots \lam_{n}}}
\\
& \hphantom{h(\alp) = \sum}
\vartht^{i_1}_{\up_1} \wed\widehat{\dots}\wed
\vartht^{i_{k-n+l}}_{\up_{k-n+l}} \wed 
d^{\lam_{1}} \wed\dots\wed d^{\lam_{n}} \,,
\end{align*}
where $0 \leq l \leq n$ and the sum is over the subsets
\beq
\{ {^{j_1}_{\uq_{1}}} \dots {^{j_l}_{\uq_{l}}} \} \sub
\{ {^{i_1}_{\up_1}} \dots {^{i_{k-n+l}}_{\up_{k-n+l}}} \} \,,
\eeq
and $\widehat{\dots}$ stands for suppressed indexes (and corresponding contact 
forms) belonging to one of the above subsets.\END
\eRm

\myskip

Now, we apply the conclusion of inclusion \eqref{wed split and ssp} of 
Preliminaries to the subsheaf $\For{k}_{r} \sub \Hor{k}_{(r+1,r)}$.  To this 
aim, we want to find the image of $\For{k}_{r}$ under the projections of the 
above splitting.

We denote the restrictions of $H,V$ to $\For{k}_{r}$ by $h,v$.

Next theorem is devoted to a characterisation of the image of $\For{k}_{r}$
under $H$.

\bTh\label{char. of h}
Let $0 < k\leq n$, and denote
\beq
\Hor{k}{_{r+1}^{h}} \byd h (\For{k}_r) \,.
\eeq

Then, we have the inclusion $\Hor{k}{_{r+1}^{h}} \sub \Hor{k}{_{r+1}^P}$.

Moreover, the sheaf $\Hor{k}{_{r+1}^{h}}$ admits the following 
characterisation: a section $\alp \in \Hor{k}{_{r+1}^P}$ is a section of the 
subsheaf $\Hor{k}{_{r+1}^h}$ if and only if there exists a section $\bet \in 
\For{k}_{r}$ such that
\beq
(j_r s)^{*}\bet = (j_{r+1}s)^{*}\alp
\eeq
for each section $s:\bX\to\bY$.
\eTh

\bPf
If $s : \bX \to \bY$ is a section, then the following identities
\beq
(j_{r}s)^*\bet = (j_{r+1}s)^*h(\bet) \,, 
\qquad\quad
(j_{r+1}s)^*v(\bet) = 0 \,,
\eeq
yield
\beq
\alp = h(\bet) \quad \Leftrightarrow \quad (j_{r}s)^*\bet = (j_{r+1}s)^*\alp
\eeq
for all $\alp \in \Hor{k}{_{r+1}^{P}}$ and $\bet \in \For{k}_{r}$.\QED
\ePf

\bRm
It comes from the above Theorem that not any section of $\Hor{k}{_{r+1}^{P}}$ 
is a section of $\Hor{k}{_{r+1}^{h}}$; indeed, a section of 
$\Hor{k}{_{r+1}^{P}}$ in general contains \lq too many monomials\rq\ with 
respect to a section of $\Hor{k}{_{r+1}^{h}}$. This can be seen by means of 
the following example. Consider a one--form $\bet \in \For{1}_{0}$.
Then we have the coordinate expressions
\beq
\bet = \bet_{\lam}d^\lam + \bet_{i}d^i \,,
\qquad\quad
h(\bet) = (\bet_{\lam} + y^i_\lam\bet_{i}) d^\lam \,.
\eeq
If $\alp \in \Hor{1}{_{1}^P}$, then we have the coordinate expression
\beq
\alp = (\alp_{\lam} + y^j_{\mu}\alp^{\mu}_{j}{_{\lam}}) d^\lam \,.
\eeq
It is evident that, in general, there does not exists 
$\bet \in \For{1}_{r}$ such that $h(\bet) = \alp$.\END
\eRm

\bCr\label{Hor aff}
Let $\dim\bX = 1$. Then we have
\beq
\Hor{1}{_{r+1}^{h}} = \Hor{1}{_{r+1}^{P}} \,.
\eeq
\eCr

\bPf
From the above coordinate expressions. See also \cite{Kru95a,Vit95}.\QED
\ePf

\bLm
The sheaf morphisms $H,V$ restrict on the sheaf $\For{k}_{r}$ to the 
surjective sheaf morphisms
\beq
h : \For{1}_{r} \to \Hor{1}{_{r+1}^h} \,,
\qquad
v : \For{1}_{r} \to \Con{1}{_r} \,.
\eeq
\eLm

\bPf
The restriction of $H$ has already been studied.  As for the restriction of
$V$, it is easy to see by means of a partition of the unity that it is 
surjective on $\Con{1}{_r}$.\QED
\ePf

\bTh\label{splitting fin.}
The splitting of proposition \ref{graded} yields the inclusion
\beq
\For{k}_r \sub \bigoplus_{l=0}^{k} \Con{k-l}{_r}\wed\Hor{l}{_{r+1}^{h}} \,,
\eeq
and the splitting projections restrict to surjective maps.
\eTh

\bPf
In fact, for any $i \leq k$ the restriction of the projection 
\beq
\Hor{k}_{(r+1,r)}  \to \Con{k-l}_{(r+1,r)}\wed\Hor{l}_{r+1}
\eeq
of the splitting of proposition \ref{graded} to the sheaf $\For{k}_{r}$
takes the form
\beq
\For{k}_{r} \to \Con{k-l}_{r}\wed\Hor{l}{_{r+1}^{h}}
\sub \Con{k-l}_{(r+1,r)}\wed\Hor{l}_{r+1} \,.
\eeq
The above inclusion can be tested in coordinates.
For the sake of simplicity, let us consider a global section
$\alp \in \Con{k-l}_{r}\wed\Hor{l}{_{r+1}^{h}}$ where $0 \leq l \leq n$.
We have the coordinate expression
\begin{align*}
& \alp = y^{j_1}_{\uq_{1}+\lam_{1}} \dots y^{j_{h}}_{\uq_{h}+\lam_{h}}
\alp  
{_{i_1 \, \dots \, i_{k-l} \, j_{1} \dots j_{h}}
^{\up_1 \dots \up_{k-l} \uq_{1} \dots \uq_{h}}}
{_{\lam_{h+1} \dots \lam_{l}}}
\\
&\hphantom{
\alp = y^{j_1}_{\uq_{1}+\lam_{1}} \dots y^{j_{h}}_{\uq_{h}+\lam_{h}}
}
\vartht^{i_{1}}_{\up_{1}} \wed\dots\wed \vartht^{i_{k-l}}_{\up_{k-l}} \wed
d^{\lam_{1}} \wed\dots\wed d^{\lam_{l}} \,,
\end{align*}
where $0 \leq |\up_{i}|,|\uq_{i}| \leq r$ and $0 \leq h \leq n$.
If $\{\psi_{i}\}$ is a partition of the unity on $\For{0}_{r}$ subordinate to 
a coordinate atlas, let
\beq
\Tilde{\alp}_{i} \byd \psi_{i} \, 
\Tilde{\alp}  
{_{t_1 \dots t_{r} }^{\us_1 \dots \us_{r}}}
{_{\lam_{r+1} \dots \lam_{k}}}
d^{t_{1}}_{\us_{1}} \wed\dots\wed d^{t_{r}}_{\up_{r}} \wed
d^{\lam_{r+1}} \wed\dots\wed d^{\lam_{k}} \,,
\eeq
where the set of pairs of indices 
$\{ {^{t_1}_{\us_{1}}} \dots {^{t_r}_{\us_{r}}} \}$
is a permutation of the set of pairs of indices 
$\{ {^{i_1}_{\up_{1}}} \dots {^{i_{k-l}}_{\up_{k-l}}} 
{^{j_1}_{\uq_{1}}} \dots {^{j_{l}}_{\uq_{l}}} \}$.
Then
\begin{enumerate}
\item $\sum_{i}\Tilde{\alp}_{i}$ is a global section of $\For{k}_r$;
\item the projection of $\sum_{i}\Tilde{\alp}_{i}$ on 
$\Con{k-l}_{r}\wed\Hor{l}{_{r+1}^{h}}$ is $\alp$.
\end{enumerate}
The proof is analogous for $k>n$.\QED
\ePf

We remark that, in general, the above inclusion is a proper inclusion:
in general, a sum of elements of the direct summands is not an element
of $\For{k}_{r}$.

\bCr\label{horizontal lift}
The sheaf morphism $H$ restricts on the sheaf $\For{k}_{r}$ to the surjective
sheaf morphisms
\begin{align*}
&h : \For{k}_r \to \Hor{k}{_{r+1}^h} \qquad k \leq n \,,
\\
&h : \For{k}_r \to \Con{k-n}{_r}\wed\Hor{n}{_{r+1}^h} \qquad k>n \,.
\end{align*}
\eCr

\section{Horizontal and vertical differential}

The derivations $i_h$, $i_v$, and the exterior differential $d$ yield two
derivations of degree one (see \cite{Sau89,Cos94}). Namely, we define
the {\em horizontal\/} and {\em vertical differential\/} to be the sheaf
morphisms
\begin{gather*}
d_h \byd i_h\com d - d\com i_h : \For{k}_r \to \For{k}_{r+1} \,,
\\
d_v\byd i_v\com d - d\com i_v : \For{k}_r \to \For{k}_{r+1} \,,
\end{gather*}

It can be proved (see \cite{Sau89}) that $d_h$ and $d_v$ fulfill the 
properties
\begin{gather*}
   d_h^2=d_v^2=0~,\qquad d_h\com d_v+d_v\com d_h=0~,
\\
   d_h+d_v=(\pi^{r+1}_r)^*\com d~,
\\ 
   (j_{r+1}s)^*\com d_v=0~,\qquad d\com (j_rs)^*=(j_{r+1}s)^*\com d_h~.
\end{gather*}

The action of $d_h$ and $d_v$ on functions $f:J_r\bY\to\R$ and one--forms 
on $J_r\bY$ uniquely characterises $d_h$ and $d_v$. We have the
coordinate expressions
\begin{gather*}
   d_hf = ({\cyrm{d}_{r+1}})_\lam .fd^\lam = 
   (\der_\lam f + y^i_{\up + \lam}\der_i^{\up}f)d^\lam \,,
\\
   d_hd^\lam = 0 \,,\qquad d_hd^i_{\up} = -d^i_{{\up}+\lam}\wed d^\lam \,,
   \qquad d_{h}\vartht^i_{\up} = -\vartht^i_{{\up}+\lam}\wed d^\lam \,,
\\
   d_vf = \der_i^{\up} f \vartht^i_{\up}~,
\\
   d_vd^\lam = 0 \,,\qquad d_vd^i_{\up}=d^i_{{\up}+\lam}\wed d^\lam \,,
   \qquad d_{v}\vartht^i_{\up} = 0 \,.
\end{gather*}
We note that 
\beq
-d^i_{{\up}+\lam}\wed d^\lam = 
-\vartht^i_{{\up}+\lam}\wed d^\lam +y^i_{{\up}+\lam +\mu} d^\mu \wed d^\lam = 
-\vartht^i_{{\up}+\lam}\wed d^\lam \,.
\eeq

Finally, next Proposition analyses the relationship of $d_h$ and $d_v$ with
the splitting of Proposition \ref{graded}.
\bPr
 \label{diffhv}
   We have
   \begin{gather*}
      d_h\left(\Hor{k}_{r}\right) \sub \Hor{k+1}{_{r+1}}\,,
      \qquad
      d_v\left(\Hor{k}_{r}\right)
      \subset\Con{1}{_r}\wed\Hor{k}_{r} \,,
   \\
      d_h\left(\Con{k}_{(r,r-1)}\wed\Hor{h}_{r}\right)
      \subset \Con{k}{_{(r+1,r)}}\wed\Hor{h+1}_{r+1} \,,
      \qquad
      d_h\left(\Con{k}_{(r,r-1)}\wed\Hor{n}_{r}\right) = \{0\} \,,
   \\  
      d_v\left(\Con{k}_{(r,r-1)}\right)\subset\Con{k+1}{_r} \,,
      \qquad
      d_v\left(\Con{k}_{r}\right)\subset\Con{k+1}{_r} \,,
   \end{gather*}
\ePr   
\bPf
   From the action of $d_h$, $d_v$ on
   functions and local coordinate bases of forms.\QED
\ePf

\chapter{Variational sequence}

In this chapter, we recall the theory of variational
sequences on finite order jet bundles, as was developed by Krupka in
\cite{Kru90}. Our main aim is to present a concise summary of the theory 
in order to introduce the reader to our notation.

Starting from the de Rham exact sheaf sequence on $J_r\bY$, we find  a
natural exact subsequence. This subsequence is not the unique exact and 
natural one that we might consider; our choice is inspired by the calculus of
variations, as it is shown in Appendix. Then we will define the 
($r$--th order) variational sequence to be the quotient of the de 
Rham  sequence on $J_r\bY$ by means of the above exact subsequence.
\myskip

We start by considering the {\em de Rham exact sequence of sheaves\/} on
$J_r\bY$
\beq
\diagramstyle[size=2.5em]
\begin{diagram}   
   0 & \rTo & \R & \rTo & \For{0}_r & \rTo^{d} & \For{1}_r &
   \rTo^{d} & \dots & \rTo^{d} & \For{J}_r &
   \rTo^{d} & 0~,
\end{diagram}
\eeq
where $J = \dim J_{r}\bY$ (see \cite{Sau89}).
\section{Contact subsequence}

We are able to provide several natural subsequences of the de Rham 
sequence. For example, natural subsequences of the de Rham sequence arise by 
considering the ideals generated in $\For{k}_r$ by its natural subsheaves
$\Hor{1}_{(r,s)}$, $\Con{1}_{(r,s)}$, $\dots$
Not all natural subsequences of the de Rham sequence turn out to be exact.
In this subsection, we study an exact natural subsequence of the de Rham 
sequence, which is of particular importance in the variational 
calculus, although being defined independently (see the Appendix).
\myskip

We introduce a new subsheaf of $\For{k}_{r}$. Namely, we set
\beq
\cC\For{k}_{r} = \{ \alp\in\For{k}_{r} \, | \, (j_{r}s)^{*}\alp = 0
\; \text{for every section}\;\, s:\bX\to\bY\} \,.
\eeq 

The definition of the above subsheaf is clearly inspired by 
the calculus of variations (see \cite{Kru90,Kru95a,Kru95b,Vit96a} and 
Appendix).

\bLm\label{cC and ker h}
We have
\begin{align*}
&\cC\For{k}_{r} = \ker h\, \quad \text{if} \quad 0\leq k \leq n \,,
\\
&\cC\For{k}_{r} = \For{k}_{r} \quad \text{if} \quad k > n \,.
\end{align*}
\eLm

\bPf
Let $\alp \in \For{k}_{r}$. Then, for any section $s:\bX\to\bY$ we have 
\beq
(j_{r}s)^{*} \alp = (j_{r+1}s)^{*} h(\alp) \,,
\eeq
and $\alp \in \ker h$ implies $\alp \in \cC\For{k}_{r}$. Conversely,
suppose $\alp \in \cC\For{k}_{r}$. Then we have
\beq
(j_{r+1}s)^{*} h(\alp) = 
h(\alp)_{\lam_{1}\dots\lam_{k}} \com j_{r+1}s \,\,\,
d^\lam_{1} \wed\dots\wed d^\lam_{k} \,,
\eeq
hence $h(\alp) = 0$.

The first assertion comes from the above identities and $\dim\bX = n$.\QED
\ePf

We define the subsheaf $\Thd{k}_r \sub \For{k}_r$ to be the sheaf generated 
by the presheaf $\ker h + d\ker h$, \ie
\beq
\Thd{k}_{r} \byd \ker h + \olin{d\ker h} \,.
\eeq
Of course, $\ker h$ is a sheaf. We recall that $\olin{d\ker h}$ 
consists of sections $\alp\in\For{k}_r$ which are of the local type
$\alp = d \bet$, with $\bet \in d\ker h$.

\bRm
If $\dim\bX = 1$ we have two important facts
\begin{enumerate}
\item $\ker h = \Con{k}_{(r,r-1)}$;
\item the above sum
turns out to be a direct sum \cite{Kru95a,Vit95}.\END
\end{enumerate}
\eRm

\bLm\label{cC and Thd}
If $0\leq k\leq n$, then $d\ker h \sub \ker h$, so that 
$\Thd{k}_r = \cC\For{k}_{r}$.
\eLm

\bPf
By the above Lemma, if $\alp \in \ker h$, then
for any section $s:\bX\to\bY$ we have $(j_{r}s)^{*}\alp = 0$, hence
$(j_{r}s)^{*}d\alp = 0$. So, $d\alp \in \ker h$.\QED
\ePf

It is clear that $\Thd{k}_r$ is a subsheaf
of $\For{k}_r$. Thus, we say the following natural subsequence
\bdg
   0 & \rTo &
   \Thd{1}_r & \rTo^{d} &
   \Thd{2}_r & \rTo^{d} &
   \dots & \rTo^{d} &
   \Thd{I}_r & \rTo^{d} & 0
\edg
to be the {\em contact subsequence} of the de Rham sequence.  We note that, in 
general, the sheaves $\Thd{k}_r$ are not the sheaves of sections of a vector 
subbundle of $T^*J_r\bY$. 

\bRm
In general, $I$ depends on the dimension of the fibers of $J_{r}\bY \to \bX$; 
its value is given in \cite{Kru90}.\END
\eRm

The following theorem is proved in \cite{Kru90} by means of a contact homotopy 
formula.

\bTh
   The contact subsequence is exact \cite{Kru90}.
\eTh

\bPr
   The sheaves $\Thd{k}_r$ are soft sheaves \cite{Kru90}.  
\ePr

\bPf
   We rephrase the proof of Krupka for convenience of the
   reader by adapting it to our notation.
   
   It can be easily seen that the sheaves $\cC\For{k}_{r}$ are soft. Let us 
   consider the short exact sequence
   \beq
      0\to\ker d\to \cC\For{1}_{r} \oset{d}{\to}\im \ d\to 0~,
   \eeq
   From the above Theorem we have $\ker d = {0}$, and this is a soft sheaf.
   Hence $\im\  d=d(\cC\For{1}_{r})$ is soft (see \cite{Wel80}).
   By induction on $k$, the equality $\ker d=\im\  d$ on $k$--forms, and 
   exactness of the sequence
   \beq
      0\to\ker d \to \cC\For{k}_{r} \oset{d}{\to} \im\  d\to 0
   \eeq   
   we obtain that each one of the sheaves $d(\cC\For{k}_{r})$ is soft.
   
   Now, let us take into account the exact sheaf sequence
   \beq
      0\to\ker d\oset{f}{\to}\cC\For{k-1}_{r}\oplus\cC\For{k}_{r}
      \oset{g}{\to}\Thd{k}_r\to 0~,
   \eeq
   where $f$, $g$ are the
   sheaf morphisms given on each tubular neighbourhood 
   ${\pi_0^1}^{-1}(\bU)$
   (with $\bU\subset\bY$ open subset) as
   \begin{gather*}
      f_{\bU} :
      (\ker d)_{\bU}\to\left(\cC\For{k-1}_{r}\oplus\cC\For{k}_{r}
      \right)_{\bU} :\alp\mto (\alp ,-d\alp )~, 
   \\
      g_{\bU} :\left(\cC\For{k-1}_{r}\oplus\cC\For{k}_{r}\right)_{\bU}\to
      (\Thd{k}_r)_{\bU} : (\alp ,\bet )\mto d\alp +\bet ~.
   \end{gather*}
   $\ker d=\im\  d$ (on $(k-1)$--forms) implies that $\ker d$ is a soft
   sheaf, and, being $\cC\For{k-1}_{r}\oplus\cC\For{k}_{r}$ soft, we 
   obtain the result.\QED
\ePf
\section{Variational bicomplex}

Here, we introduce a bicomplex by quotienting the de Rham sequence on $J_r\bY$
by the contact subsequence. We obtain a new sequence, 
the variational sequence, which turns out to be exact. In the last part of the 
section, we describe the relationships between bicomplexes on jet 
spaces of different orders.
\bPr
   The following diagram
\newdiagramgrid{Krupka}%
{1,.5,1,.5,1,.5,1,.8,1,.5,.5,1,1,.5,1,.7,.5,.5,1}
{.9,.9,.9,.9,.9,.9,.9,.9}
\beq
\diagramstyle[size=2.3em]
\begin{diagram}[grid=Krupka]
&& 0 && 0 && 0 && 0 &&&& 0 && 0  &&&&
\\
&& \dTo && \dTo && \dTo && \dTo &&&& \dTo && \dTo &&&&
\\
0 & \rTo & 0 & \rTo & 0 & \rTo &
\Thd{1}_r & \rTo^d & \Thd{2}_r & \rTo^d & \dots &
\rTo^d & \Thd{I}_r & \rTo^d & 0 & \rTo & \dots & \rTo & 0
\\
&& \dTo && \dTo && \dTo && \dTo &&&& \dTo && \dTo &&&&
\\
0 & \rTo & \R & \rTo & \For{0}_r & \rTo^d &
\For{1}_r & \rTo^d & \For{2}_r & \rTo^d & \dots & \rTo^d &
\For{I}_r & \rTo^d & \For{I+1}_r & \rTo^d & \dots & \rTo^d & 0
\\
&& \dTo && \dTo && \dTo && \dTo &&&& \dTo && \dTo &&&&
\\
0 & \rTo & \R & \rTo & \For{0}_r & \rTo^{\cE_{0}} &
\For{1}_r/\Thd{1}_r & \rTo^{\cE_{1}} & \For{2}_r/\Thd{2}_r & 
\rTo^{\cE_{2}} &
\dots & \rTo^{\cE_{I-1}} & \For{I}_r/\Thd{I}_r & \rTo^{\cE_{I}} &
\For{I+1}_r & \rTo^{d} & \dots & \rTo^{d} & 0
\\
&& \dTo && \dTo && \dTo && \dTo &&&& \dTo && \dTo &&&&
\\
&& 0 && 0 && 0 && 0 &&&& 0 && 0  &&&&
\end{diagram}
\eeq
   is a commutative diagram, where rows and columns are exact.
\ePr

\bPf
We have to prove only the exactness of the bottom row of the diagram. 
But this follows from the exactness of the other rows and of the
columns.\QED
\ePf

\bDf
   The above diagram is said to be the $r$--th order {\em 
   variational bicomplex\/} associated with the fibred manifold $\bY\to\bX$
   (see \cite{Kru90}). 
   
   We say the bottom row of the above diagram to be the $r$--th order
   {\em variational sequence\/} associated with the fibred manifold
   $\bY\to\bX$.\END
\eDf

\bPr
   The sheaves $\For{k}_r/\Thd{k}_r$ are soft sheaves (see \cite{Kru90}).
\ePr

\bPf
In fact, each column is a short exact sheaf sequence in which 
$\Thd{k}_r$ and $\For{k}_r$ are soft sheaves (see \cite{Wel80}).\QED
\ePf 

\bCr
The variational sequence is a soft resolution of the constant sheaf
$\R$ over $\bY$ \cite{Kru90}.
\eCr

\bPf
In fact, except $\R$, each one of the sheaves in the sequence is 
soft \cite{Wel80}.\QED
\ePf

The most interesting consequence of the above corollary is the following 
one (for a proof, see \cite{Wel80}). Let us consider the cochain complex
\beq
\diagramstyle[size=2.5em]
\begin{diagram}
   0 & \rTo &
   \R_{\bY} & \rTo &
   \left(\For{0}_r\right)_{\bY} & \rTo^{d} &
   \left(\For{1}_r/\Thd{1}_r\right)_{\bY} & \rTo^{\cE_1} &
   \dots & \rTo^{d} &
   \left(\For{J}_r\right)_{\bY} & \rTo^{d} & 0
\end{diagram}
\eeq
and denote by $H^k_{\text{VS}}$ the $k^{\text{th}}$--cohomology group of 
the above cochain complex.
\bCr
   For all $k\geq 0$ there is a natural isomorphism
   \beq
      H^k_{\text{VS}}\simeq H^k_{\text{de Rham}}\bY
   \eeq
   (see \cite{Kru90}).
\eCr
\bPf
   In fact, the variational sequence is a soft
   resolution of $\R$, hence the cohomology of the sheaf $\R$ is
   naturally isomorphic to the cohomology of the above cochain complex.
   Also, the de Rham sequence gives rise to a cochain complex of global 
   sections, whose cohomology is naturally isomorphic to the cohomology of the
   sheaf $\R$ on $\bY$. Hence, we have the result by a composition of
   isomorphisms. (See \cite{Wel80} for more details on the above natural
   isomorphisms.)\QED
\ePf

\myskip

Finally, we investigate the relationship between variational bicomplexes of
different orders. To this purpose, we recall the intrinsic inclusions 
($0\leq s\leq r$)
\beq
   \For{k}_s\simeq {\pi^{r}_s}^*\For{k}_s\subset\For{k}_{r}~,\qquad
   \Thd{k}_s\simeq {\pi^{r}_s}^*\Thd{k}_s\subset\Thd{k}_{r}~,
\eeq
and the isomorphism
\beq
   \left(\For{k}_s/\Thd{k}_s\right)\simeq
   \left( {\pi^{r}_s}^*\For{k}_s/{\pi^{r}_s}^*\Thd{k}_s\right)~.
\eeq
\bLm
   Let $s\leq r$. Then, the above inclusions induce the injective sheaf
   morphism (see \cite{Kru90})
   \beq
      \chi^{r}_s :\left(\For{k}_s/\Thd{k}_s\right)\to
      \left(\For{k}_{r}/\Thd{k}_{r}\right) : 
      [\alp ] \mto [{\pi^{r}_s}^*\alp ]~,
   \eeq
   where $[\alp ]$ denotes an equivalence class of a form on $J_s\bY$.
\eLm
\bPf
   The above morphism $\chi^{r}_s$ is well--defined, because
   \beq
      [\alp ] = [\bet ] \Rightarrow [{\pi^{r}_s}^*\alp ]
      = [{\pi^{r}_s}^*\bet ]
   \eeq
   due to the above inclusions.
   
   The morphism is injective too. For if $\alp\in\For{k}_{s}$ and
   $\bet\in\For{k}_s$ such that
   \beq
      [{\pi^{r}_s}^*\alp ] = [{\pi^{r}_s}^*\bet ] ~,
   \eeq
   then, being ${\pi^{r}_s}^*(\alp-\bet )\in {\pi^{r}_s}^*\For{k}_s$,
   and ${\pi^{r}_s}^*(\alp-\bet ) \in \Thd{k}_{r}$, it must be
   ${\pi^{r}_s}^*(\alp-\bet )\in {\pi^{r}_s}^*\Thd{k}_s$, hence
   $[\alp ] = [\bet ]$.\QED
\ePf

\bRm\label{filtration}
It is clear that, if $t\leq s\leq r$, then $\chi^r_s\com\chi^s_t=\chi^r_t$.

We have the commutative diagrams
\bdg
\For{k}_{r} & \rTo^{d} & \For{k+1}_{r} &&
\Thd{k}_{r} & \rTo^{d} & \Thd{k+1}_{r}
\\
\uTo_{{\pi^{r}_s}^*} & & \uTo^{{\pi^{r}_s}^*} &&
\uTo_{{\pi^{r}_s}^*} & & \uTo^{{\pi^{r}_s}^*} &&
\\
\For{k}_s & \rTo^{d} & \For{k+1}_s  &&
\Thd{k}_s & \rTo^{d} & \Thd{k+1}_s
\edg
hence the following commutative diagram holds
\bdg
\For{k}_{r}/\Thd{k}_{r} & \rTo^{\cE_k} & \For{k+1}_{r}/\Thd{k+1}_{r}
\\
\uTo_{{\pi^{r}_s}^*} & & \uTo^{{\pi^{r}_s}^*}
\\ 
\For{k}_s/\Thd{k}_s & \rTo^{\cE_k} & \For{k+1}_s/\Thd{k+1}_s 
\edg

We can summarise the above commutative diagrams stating the existence of a
three--dimensional commutative diagram (which is not exact), whose
bi--dimensional slices are the variational bicomplexes of order $1$, $2$,
$\dots$.\END
\eRm
\chapter{Representation of the variational sequence}

In this section we find suitable sheaves of fibred morphisms that are
isomorphic to the quotient sheaves of the variational
sequence.

As a consequence, we will recover the sheaves of the geometric  objects that
arise in the variational calculus (like Lagrangians, Euler--Lagrange
morphisms,\dots ). Also, we will be able to give 
an intrinsic formulation of the Helmholtz conditions of local variationality.
By the way, one can see that in the infinite--jet formalism one loses
information relatively to the order of the jet in which objects really
\lq live\rq .
\myskip

We start by restricting our analysis to the following {\em short exact
subcomplex}
\newdiagramgrid{short}%
{1,.5,1,.5,1,.8,1,.8,1,.8,1,1,1,1,1}
{.9,.9,.9,.9,.9,.9,.9,.9}
\beq
\diagramstyle[size=2.3em]
\begin{diagram}[grid=short]
&& 0 && 0 && 0 &&&& 0 && 0 &&&&
\\
&& \dTo && \dTo && \dTo &&&& \dTo && \dTo &&&&
\\
0 & \rTo & 0 & \rTo & 0 & \rTo &
\Thd{1}_r & \rTo^d & \dots &
\rTo^d & \Thd{n+1}_r & \rTo^d & d\Thd{n+1}_r & 0
\\
&& \dTo && \dTo && \dTo &&&& \dTo && \dTo &&&&
\\
0 & \rTo & \R & \rTo & \For{0}_r & \rTo^d &
\For{1}_r & \rTo^d & \dots & \rTo^d &
\For{n+1}_r & \rTo^d & d\For{n+1}_r & \rTo^d & 0
\\
&& \dTo && \dTo && \dTo &&&& \dTo && \dTo &&&&
\\
0 & \rTo & \R & \rTo & \For{0}_r & \rTo^{\cE_{0}} &
\For{1}_r/\Thd{1}_r & \rTo^{\cE_{1}} & 
\dots & \rTo^{\cE_{n}} & \cE_n(\For{I}_r/\Thd{I}_r) & \rTo^{\cE_{n+1}} & 0
\\
&& \dTo && \dTo && \dTo &&&& \dTo && \dTo &&&&
\\
&& 0 && 0 && 0 &&&& 0 && 0 &&&&
\end{diagram}
\eeq
due to the fact that, to our knowledge, if $k\geq n+3$, there is no 
interpretation of the $k^{\text{th}}$--column of the variational bicomplex
in terms of geometric objects of the variational calculus. We say the bottom
row of the above bicomplex to be the {\em short variational sequence}.

\section{Lagrangian}

In this section, we show that the quotient sheaves
\beq
\For{1}_r/\Thd{1}_r \,, \dots \,, \For{n}_r/\Thd{n}_r
\eeq
are isomorphic to certain subsheaves of sheaves of sections of a vector
bundle. In this way, we are able to find an explicit expression for
the sheaf morphisms $\cE_0,\dots \cE_{n-1}$.

\myskip

\bTh
Let $k\leq n$. Then, the sheaf morphism $h$ yields the isomorphism 
\beq
I_{k} : \For{k}_r/\Thd{k}_r \to \Hor{k}{_{r+1}^{h}} :
[\alp] \mto h(\alp) \,.
\eeq
\eTh

\bPf
This is by the fact that, if $k\leq n$, then $\Thd{k}_r = \ker h$,
and to the characterisation of the image of $h$ of Theorem 
\ref{char. of h}.\QED
\ePf

\bCr
The sheaf morphisms $\cE_{0}, \dots ,\cE_{n-1}$ are expressed through the 
above isomorphisms $I_{k}$ as
\beq
\cE_{k}(h(\alp)) = \cE_{k}(h(d\alp)) \,.
\eeq
\eCr

As an example, we have $\cE_{0} = d_{h}$. It is easy to compute coordinate 
expressions for $\cE_{0}, \dots ,\cE_{n-1}$ via the above Corollary.

\bDf
Let us set $\Var{k}_r\byd\Hor{k}{_{r+1}^h}$.  

We say a section $L\in\Var{n}_r$ to be a {\em $r$--th order generalised 
Lagrangian}.\END
\eDf

It is worth to note that the sheaf of the $r$--th order Lagrangians of the 
standard literature is $\Hor{n}_r$, and that 
$\Hor{n}_r\subset\Hor{n}{_{r+1}^h}$ (see the Appendix).

\section{Euler--Lagrange morphism}

In this section we will show that the quotient sheaf $\For{n+1}_r/\Thd{n+1}_r$ 
of the variational sequence is isomorphic to certain subsheaves of sheaves of 
sections of a vector bundle.  In this way, we are able to find an explicit 
coordinate expression for the sheaf morphism $\cE_n$.

Throughout this chapter we will adopt the notation
\beq
\olin{d_{h}}(\Con{k-n}_{r}\wed\Hor{n-1}{_{r+1}^h}) \byd
\olin{d_{h}(\Con{k-n}_{r}\wed\Hor{n-1}{_{r+1}^h})} \,,
\eeq
for evident practical reasons.

It is possible to introduce a first simplification of the quotient
sheaves.

\bLm\label{cC and dh}
If $k>n$, then the restriction of $h$ to the sheaf $\Thd{k}_r$ 
is the surjective presheaf morphism 
\beq
h : \Thd{k}_r \to h(\olin{d\ker h}) \,.
\eeq
Moreover, pull--back yields the natural inclusion
\beq
h(\olin{d\ker h})  \sub \olin{d_{h}}(\Con{k-n}_{r}\wed\Hor{n-1}{_{r+1}^h}) 
\sub h(\Thd{k}_{r+1}) = h(\olin{d\ker h}) \,,
\eeq
which turns out to be an equality if $\dim \bX = 1$.
\eLm

\bPf
The first statement is obvious. We have the natural 
identification $\ker h \simeq v(\ker h)$, which yields
\beq
h(d\ker h) \simeq 
h(d_{h}v(\ker h) + d_{v}v(\ker h)) \simeq
hd_{h}v(\ker h) \,,
\eeq
due to Proposition \ref{diffhv}. The same Proposition yields the inclusion
\beq
hd_{h}v(\ker h) \sub d_{h}(\Con{k-n}_{r}\wed\Hor{n-1}{_{r+1}^h}) 
\,,
\eeq
hence the inclusions of the statement.

If $\dim \bX = 1$, then $\ker h = \Con{k}_{(r,r-1)}$
\cite{Kru95a,Vit95}, hence the result.\QED
\ePf

\bPr\label{first isomorphism}
Let $k>n$. Then, the projection $h$ induces the natural sheaf isomorphism
\beq
\left(\For{k}_r/\Thd{k}_r\right)\to
\left(\Con{k-n}{_r}\wed\Hor{n}{_{r+1}^h}\right) \big / h(\olin{d\ker h})
: [\alp ]\mto [h(\alp )]~.
\eeq
\ePr

\bPf
The map is clearly well defined.
   
Also, the map is injective, for if $\alp , \alp '\in\For{k}_r$, then
\beq
[h(\alp )] = [h(\alp ')] \Rightarrow h(\alp -\alp ')=hdp~,
\eeq
with $p \in \ker h$. Hence
\beq
\alp -\alp ' = v(\alp -\alp '-dp) + dp \,,
\eeq
where, being $dp \in \For{k}_{r}$ and $\alp -\alp ' \in \For{k}_{r}$, we have
$v(\alp -\alp '-dp) \in \For{k}_{r}$. Due to $h\circ v = 0$, we have
$[\alp - \alp'] = 0$.
   
Finally, the map is surjective, due to the surjectivity of $h$.\QED
\ePf

\bRm
In spite of the apparent complexity of the quotient sheaf 
\newline
$\left(\Con{k-n}{_r}\wed\Hor{n}{_{r+1}^h}\right) 
\big /h(\olin{d\ker h})$,
we notice that it is made with proper subsheaves of the sheaves
$\For{k}_{r+1}$ and $\Thd{k}_{r+1}$. Hence, our search for a natural 
representative in each equivalence class will be considerably 
simplified.\END
\eRm

\bRm\label{higher order rep.}
Let $0\leq s\leq r$. Then, the sheaf injection $\chi^r_s$ induces
the sheaf injection
\beq
\left(\Con{k-n}{_s}\wed\Hor{n}{_{s+1}^h}\right)\big /
h(\olin{d\ker h})\to
\left(\Con{k-n}{_r}\wed\Hor{n}{_{r+1}^h}\right)\big /
h(\olin{d\ker h}) ~.\END
\eeq
\eRm

\myskip

As for the sheaf $\For{n+1}_r/\Thd{n+1}_r$, taking into account the 
isomorphism of Proposition \ref{first isomorphism} and the identification
of Lemma \ref{cC and dh}, we have two main tasks:
\begin{enumerate}
\item to find for all
$\alp\in\Con{1}{_r}\wed\Hor{n}{_{r+1}^h}$ a natural (and possibly unique, in 
some sense) $n$--form $F_{\alp} \in h(\olin{d\ker h})$ in such a way that 
the sheaf morphism
\beq
I_{n+1} : \left(\Con{1}{_r}\wed\Hor{n}{_{r+1}^h}\right) \big /
h(\olin{d\ker h}) \to \For{n+1}_{r+s}
: [\alp ]\mto \alp +F_{\alp}
\eeq
is injective (for some $s\in\BN$);
\item to characterise the image of the above sheaf morphism, so to obtain
a sheaf of sections of a vector bundle that is isomorphic to
$\For{n+1}_r/\Thd{n+1}_r$.
\end{enumerate}

\myskip

The above first problem can be solved by means of a result by Kol\' a\v r 
\cite{Kol83}. To proceed further, we need some notation.  On the domain of 
any chart, we set
\begin{gather*}
\ome \byd \sum_{\lam_{1}, \dots ,\lam_{n} = 1}^{n} 
d^{\lam_{1}} \wed\dots\wed d^{\lam_{n}} = n! d^{1} \wed\dots\wed d^n \,,
\\
\ome_{\lam} \byd \, i_{\der_{\lam}} \ome \,,
\qquad
\ome_{\lam\mu} \byd i_{\der_{\mu}} (\ome_{\lam}) \,.
\end{gather*}
We have 
\beq
\sum_{\mu = 1}^n d^\mu\wed\ome_{\lam} = \ome \,.
\eeq 
If $\bU\subset\bY$ is a coordinate open subset and 
$f\in\left(\For{0}_r\right)_{\bU}$, then we set, by induction
\beq
J_{\lam}f\byd (\cyrm{d}_{r+1})_{\lam} f \,, \qquad
J_{\up+\lam}f\byd J_{\lam} J_{\up} f \,;
\eeq
analogously, we denote by $L_{J_{\up}}$ the iterated Lie derivative.
We have the characterisation
\beq
J_{\up}f \com j_{r+|\up|} s = \der_{\up}(f \com j_{r}s) \,,
\eeq
A Leibnitz' rule holds (see \cite{Sau89}); if 
$g\in\left(\For{0}_r\right)_{\bU}$, then we have
\beq
J_{\up}(fg)=\sum_{\uq + \ut = \up} 
\, \frac{\up !}{\uq ! \ut !} \, J_{\uq}f \, J_{\ut}g \,.
\eeq
Let $u:\bY\to V\bY$ be a vertical vector field with coordinate expression
$u=u^i\der_i$. Then, the coordinate expression of the prolongation 
$u_r:J_r\bY\to VJ_r\bY$ is $u_r=J_{\up}u^i\der_i^{\up}$.

\bTh\label{higher-order Lag.}
(First variation formula for higher--order variational calculus
\newline\cite{Kol83}) Let $\alp\in\For{1}_r\wed\Hor{n}_r$ $\simeq$ $\Con{1}_{r}
\wed \Hor{n}_{r}$.  Then there is a unique pair of sheaf morphisms
\beq
E_{\alp} \in \Con{1}_{(2r,0)}\wed\Hor{n}{_{2r}} \,,
\qquad
F_{\alp} \in \Con{1}_{(2r,r)} \wed \Hor{n}{_{2r}} \,,
\eeq
such that 

i. $(\pi^{2r}_{r})^*\alp=E_{\alp}-F_{\alp}$;

ii. $F_\alp$ is locally of the form $F_{\alp} = d_{h}p_{\alp}$, with $p_{\alp}
\in \Con{1}_{(2r-1,r-1)}\wed\Hor{n}{_{2r}}$.
\eTh

\bPf
The proof is carried on by induction. We set, in a coordinate neighbourhood,
\beq
\alp = \alp^{\up}_{i} \, \vartht^i_{\up}\wed\ome \,,
\qquad
p_{\alp} = p{^{\uq}_{i}}{^\lam} \, \vartht^i_{\uq}\wed\ome_{\lam} \,,
\qquad
E_{\alp} = E_i \, \vartht^i\wed\ome \,,
\eeq
where $0\leq |\up | \leq r$ and $0\leq |\uq |\leq r-1$,
hence we have
\beq
d_{h}p_{\alp} = -J_{\lam} p{^{\uq}_{i}}{^\lam} \, \vartht^i_{\uq}\wed\ome
 - p{^{\uq}_{i}}{^\lam} \, \vartht^i_{\uq + \lam}\wed\ome \,.
\eeq

The requirements on $E_{\alp}$ and on $p_{\alp}$ yield the vanishing of 
some components of the sum
$\alp + d_{h}p_{\alp}$, hence a system of linear equations which has a unique 
pair of local solutions $E_{\alp}$ and $d_{h}p_{\alp}$. 
In particular, we have
\bEq\label{EL coord.}
E_{\alp}=(-1)^{|\up |}J_{\up }\alp_i^{\up} \, \vartht^i\wed\ome \,,
\qquad
0 \leq |\up | \leq r \,.
\eEq
The uniqueness ensures that $E_{\alp}$ and $d_{h}p_{\alp}$ are intrinsically 
characterised, hence they yield two sections $E_{\alp}$ and $F_{\alp}$
which fulfill the requirement of the statement; in particular, they have the 
same domain of definition as $\alp$.\QED
\ePf

\bRm\label{uniqueness of p}
As it is proved in \cite{Kol83}, to any $F_{\alp}$ there always exists 
a section $p_\alp \in  \Con{1}_{(2r-1,r-1)}\wed\Hor{n}{_{2r}}$ (with the same 
domain as $\alp$) such that $F_{\alp}$. But in general such a $p_\alp$ is not 
unique. In fact \cite{Kol83}, by adding to $p_{\alp}$ the 
horizontal differential of a suitable $n-1$--form we obtain another form which 
fulfills the conditions of the statement. Anyway, we have some particular 
cases where a form $p_{\alp}$ can be uniquely determined.
\begin{enumerate}
\item Suppose that $\dim \bX = 1$. Then, it can be easily proved 
\cite{AnTh92,Kru95a,Vit95} that $d_{h}p_{\alp} = 0$ implies $p_{\alp} = 0$,
so that $p_{\alp}$ is uniquely determined.
\item Suppose that $r = 1$. Then, one can easily realise that there does not 
exists a $n-1$--form such that its horizontal differential is a section of
$\Con{1}_{(2r-1,r-1)} \wed \Hor{n-1}_{2r-1}$, so that $p_{\alp}$ is uniquely 
determined. In this case, we can say even more. In fact, we are able to 
determine $p_{\alp}$ from $\alp$ by means of the natural sheaf morphism
\beq
p: \Con{1}_{1}\wed\Hor{n}_1 \to \Con{1}_{(1,0)}\wed\Hor{n-1}_{1} ~.
\eeq
which fulfills
\beq
p(d_{h}\bet )=-\bet 
\qquad\quad
\forall \bet \in \Con{1}_{(1,0)}\wed\Hor{n-1}_{1} ~.
\eeq
The above morphism was introduced in different forms by several authors
\cite{Kol93,MaMo83b,Sau89,Vit95}. If
\beq 
\alp = \alp_{i} \, d^i\wed\ome + \alp_{i}^{\lam} \, d^i_{\lam}\wed\ome \,,
\eeq
then we have the coordinate expression
\beq
p(\alp) = \alp_{i}^{\lam} \, \vartht^i\wed\ome_{\lam} \,.
\eeq
\item In the case $r = 2$ we are able to characterise a unique $p_{\alp}$ by 
means of an additional requirement. There is a natural morphism
\beq
s : \Con{1}_{(3,1)} \wed \Hor{n-1}_{3} \to \Con{1}_{(3,0)} \wed \Hor{n-2}_{3}
\eeq
where, if $p \in \Con{1}_{(3,1)} \wed \Hor{n-1}_{3}$ has the coordinate 
expression
\beq
p = p_{i}{^{\mu}} \, \vartht^i\wed\ome_{\mu}
+ p{_{i}^{\lam}}{^{\mu}} \, \vartht^i_{\lam}\wed\ome_{\mu} \,,
\eeq
then we have
\beq
s(p) = p{_{i}^{\lam}}{^{\mu}} \, \vartht^i\wed\ome_{\mu\lam} \,.
\eeq
It is easily proved that there exists a unique morphism 
$p_{\alp} \in \Con{1}_{(3,1)} \wed \Hor{n-1}_{3}$ such that $s(p_{\alp}) = 0$.
Such a morphism is called {\em quasisymmetric}. This result has been shown in
\cite{Kol83}. In particular, if we have the coordinate expression
\beq
\alp = \alp_{i} \, \vartht^i\wed\ome +
\alp_{i}^{\lam} \, \vartht^i_{\lam}\wed\ome +
\alp_{i}^{\lam + \mu} \, \vartht^i_{\lam + \mu}\wed\ome \,,
\eeq
then we have
\beq
p_{\alp} = 
(\alp_{i}^{\lam} - J_{\mu}\alp{_{i}^{\mu +\lam}}) \, \vartht^i\wed\ome_{\lam}
+ \alp{_{i}^{\mu +\lam}} \, \vartht^i_{\mu}\wed\ome_{\lam} \,.
\eeq
\item In the case $r \geq 3$ we have no natural ways to select a form 
$p_{\alp}$. In \cite{Kol83} a sufficient condition to the uniqueness of 
$p_{\alp}$ is given. Namely, let us denote with $T^{*}_{s}\bX$ the 
$s$--th order cotangent bundle. Let $\Gam : T^{*}\bX \to T^{*}_{r-1}\bX$
be a section which is also a linear morphism.  Then, there exists a unique 
form $p_{\alp}[\Gam]$.  Note that, if $r=2$, then we have the natural choice 
$\Gam = \id$, which yields the natural form of the previous item.
\end{enumerate}\END
\eRm

\bRm
The choice of the subsheaf 
\beq
\Con{1}_{(2r,0)}\wed\Hor{n}_{2r} \sub 
\Con{1}_{(2r,2r-1)}\wed\Hor{n}_{2r}
\eeq
will provide representatives $E_{\alp}$ of sections of the quotient sheaf 
$\For{2}_1/\Thd{2}_1$ with a minimal number of components.\END
\eRm

\bRm\label{weighted deg.}
The section $E_{\alp} \in \Con{1}_{(2r,0)}\wed\Hor{n}_{2r}$ has a peculiar 
structure with respect to the derivative coordinates of order greater than 
$r$. In fact, if we assign to the variables $y^i_{\up}$ with $| \up | = r+s$
the weight $s$, then it is easily seen that $E_{\alp}$ is a polynomial with 
weighted degree $r$ with respect to $y^i_{\up}$, with $| \up | > r$.
This kind of structure was first introduced and studied in \cite{KoMo90}.\END
\eRm

\bCr\label{order of E}
Let $\alp\in\Con{1}_r\wed\Hor{n}{_{r+1}^h} \sub 
\For{1}_{r+1}\wed\Hor{n}_{r+1}$. 
Then $E_{\alp}$ and $p_{\alp}$ 
are sections of the following subsheaves
\beq
E_{\alp} \in \Con{1}_{(2r,0)}\wed\Hor{n}{_{2r+1}^{h}} \,,
\qquad 
p_{\alp} \in \Con{1}_{(2r,r-1)} \wed \Hor{n-1}{_{2r}^h} \,.
\eeq
\eCr

\bPf
This depends on the form of the system 
\beq
\alp = E_{\alp} - d_{h}p_{\alp} \,,
\eeq
and on the fact that the form $\alp$ takes values into the vector bundle
$T^{*}J_{r}\bY \wed T^{*}\bX$, even if it depends on $J_{r+1}\bY$.\QED
\ePf

\bRm
In the case $\alp\in\Con{1}_r\wed\Hor{n}{_{r+1}^h}$, the section
$E_{\alp}$ has an additional feature with respect to the polynomial structure 
of Remark \ref{weighted deg.}. In fact, the coefficients of the polynomial are
polynomials of (standard) degree $n$ with respect to the coordinates 
$y^i_{\up}$, with $| \up | = r+1$.\END
\eRm

\bTh\label{character. of p}
Let $q \in \Con{1}_{(2r-1,r-1)} \wed \Hor{n-1}_{2r-1}$. Then we have
\beq
d_{h}p_{d_{h}q} = -d_{h}q \,,
\eeq
hence $E_{d_{h}q} = 0$.
\eTh

\bPf
In fact, by recalling the proof of the above Theorem, we find that
the system
\beq
d_{h}q = E_{d_{h}q} - d_{h}p_{d_{h}q}
\eeq
has the unique solutions $d_{h}p_{d_{h}q} = -d_{h}q$ and $E_{d_{h}q} = 0$.\QED
\ePf

\bRm
The above Theorem is the geometric interpretation of the well--known fact 
that \lq the Euler--Lagrange morphism annihilates divergencies\rq\ (see also
\newline\cite{Tra96}).\END
\eRm

\bPr
The sheaf morphism
\beq
\Con{1}{_r}\wed\Hor{n}{_{r+1}^h} \to \For{n+1}_{2r+1}
 :\alp \mto \alp + F_{\alp}
\eeq
induces the injective sheaf morphism
\beq
I_{n+1} :\left(\Con{1}{_r}\wed\Hor{n}{_{r+1}^h}\right) \big /
h(\olin{d\ker h})\to\For{n+1}_{2r+1}
: [\alp ] \mto \alp +F_{\alp}
\eeq
\ePr

\bPf
We make use of the injective morphism $\chi^s_{r}$ of Remark 
\ref{higher order rep.} and of Lemma \ref{cC and dh}.
The morphism $I_{n+1}$ is well--defined, due to Corollary \ref{order of E} 
and to the fact that, if
$\alp ,\bet \in \Con{1}{_r}\wed\Hor{n}{_{r+1}^h}$ such that
$\bet =\alp + F$, where $F$ is of the local form $F = d_hq$ with 
$d_hq \in d\ker h$ then
\beq
\bet + F_{\bet} =\alp + F + F_{\alp} + F_{F} \,,
\eeq
where $F_{F}=-F$ by the uniqueness in Theorem \ref{higher-order Lag.}.
   
We have to prove that the morphism is injective. Suppose that
\beq
\bet + F_{\bet}=\alp + F_{\alp}~.
\eeq
Hence $\bet -\alp = F_{\alp} - F_{\bet}$, so $[\bet - \alp] = 0$.\QED
\ePf

The final step is to characterise the image of $I_{n+1}$. 

\bTh
We have the sheaf isomorphism
\beq
I_{n+1} : \For{n+1}_r/\Thd{n+1}_r\to\Var{n+1}_r,
\eeq
where
\beq
\Var{n+1}_r \byd 
\left(\Con{1}{_r}\wed\Hor{n}{_{r+1}^h} +
\olin{d_h} (\Con{1}_{(2r,r-1)}\wed\Hor{n-1}_{2r}) \right) \cap 
\left(\Con{1}_{(2r+1,0)}\wed\Hor{n}_{2r+1}\right) ~.
\eeq
\eTh

\bPf
It comes from the isomorphism of Proposition \ref{first isomorphism},
the injective morphism $I_{n+1}$ and the characterisation of the 
image of $I_{n+1}$ provided by Theorem \ref{higher-order Lag.}.\QED
\ePf

Now, we can evaluate $\cE_{n}$ by means of the isomorphisms $I_{n},I_{n+1}$.

\bTh
Let $\alp\in\Var{n}_{r}$. Then, $\cE_{n}(\alp) \in \Var{n+1}_{r}$ coincides 
with the standard higher--order Euler--Lagrange morphism 
\cite{Fer83,FeFr82,GaMu82,Kol83,Kru83,Sau89} associated with the
generalised $r$--th order Lagrangian $\alp$, regarded as 
a standard $r+1$--th order Lagrangian.
\eTh

\bPf
In fact, Theorem \ref{higher-order Lag.} yields the standard higher--order
Euler--Lagrange morphism. Moreover, we have the inclusions
\beq
\Var{n}_{r} \sub \Hor{n}_{r+1} \sub \Var{n}_{r+1}\,.
\eeq
The result now is immediate, due to the commutativity of the inclusion of the 
bicomplex of order $r$ into the bicomplex of order $r+1$ (Remark 
\ref{filtration}).\QED
\ePf

\bDf
Let $\alp \in \For{n+1}_{r}$.

We say $E_{h(\alp)}\in\Var{n+1}_{r}$ to be the 
{\em generalised $r$--th order Euler--Lagrange morphism} associated with 
$\alp$. 

We say $p_{h(\alp)}$ to be a {\em generalised $r$ th order momentum\/} 
associated with $E_{h(\alp)}$.

We say $\cE_{n}$ to be the {\em generalised $r$--th order Euler--Lagrange 
operator}.\END
\eDf

\bRm
It is of fundamental importance to note that some theories which are based 
upon polynomial $(r+1)$--th order horizontal Lagrangians can be seen also as 
$r$--th order theories using a non--horizontal Lagrangian (see Appendix).  And 
it is worth to point out that most of second--order horizontal Lagrangians 
known in physics are affine.\END
\eRm
\section{Helmholtz morphism}

In this section we will devote ourselves to a description of 
the presheaf 
\beq
\cE_{n+1}\left(\Var{n+1}_r\right)\simeq
\cE_{n+1}\left(\For{n+1}_r/\Thd{n+1}_r\right) =
\left(d\For{n+1}_r/d\Thd{n+1}_r\right) ~. 
\eeq

In particular, we will find an isomorphism of this presheaf with a 
subpresheaf of a sheaf of sections of a vector bundle.  Hence, we will 
be able to provide an explicit expression for the map $\cE_{n+1}$.  
This will yield an intrinsic geometric object whose vanishing is 
equivalent to the Helmholtz conditions of local variationality.

\myskip

Let $E\in\Var{n+1}_{r}$. In order to evaluate the expression of 
$\cE_{n+1}(E)$, it is very difficult to find a $n+1$--form 
$\alp\in\For{n+1}_{s}$ such that $I_{n+1}([h(\alp)])=E$. So, it is 
difficult in concrete cases to use the commutativity of the 
diagram in order to compute $\cE_{n+1}$.

Hence, the most convenient way to reach our task is to
use Theorem \ref{first isomorphism} together with the isomorphism
$\For{n+1}_r/\Thd{n+1}_r \to \Var{n+1}_r$ in order to simplify the analysis
of the sheaf $\cE_{n+1}\left(\For{n+1}_r/\Thd{n+1}_r\right)$.

\bLm\label{Helm. equiv.}
We have the natural injection
\beq
\left(d\For{n+1}_r/d\Thd{n+1}_r\right) \to
\left(\Con{2}_{2r+1}\wed\Hor{n}{_{2r+2}^h}\right) \big /
h(\olin{d\ker h})
: [d\alp]\mto [dE_{h(\alp)}]~.
\eeq
\eLm
\bPf
It is a direct consequence of the decomposition
\beq
\alp =E_{h(\alp)} - d_hp_{h(\alp)} + v(\alp ) \,.
\eeq
together with $dd_h=-d_hd_v$.\QED
\ePf

Hence, we search for natural representatives of the
classes of the image of
\beq
d\Var{n+1}_r \sub
\For{1}_{2r+1}\wed\Con{1}_{(2r+1,0)}\wed\Hor{n}_{2r+1} \simeq
\Con{1}_{2r+1}\wed\Con{1}_{(2r+1,0)}\wed\Hor{n}_{2r+1} \sub
\Con{2}_{2r+1}\wed\Hor{n}{_{2r+2}^h}
\eeq
into the quotient $\left(\Con{2}_{2r+1}\wed\Hor{n}{_{2r+2}^h}\right) \big / 
h(\olin{d\ker h})$; we denote this image by $[d\Var{n+1}_r]$.  

Our task is the following one: {\em to characterise a unique representative of 
every equivalence class of $[d\Var{n+1}_r]$ by means of the higher--order 
Euler--Lagrange morphism.\/}

First of all, we need a technical Lemma.

\bLm
Let $\bet \in \Con{1}_{s}\wed\Con{1}_{(s,0)}\wed\Hor{n}_{s}$. 
Suppose that the coordinate expression of $\bet$ is
\beq
\bet = \bet{^{\up}_{i}}{_j}\vartht^i_{\up}\wed\vartht^j\wed \ome \,,
\qquad
0 \leq |\up | \leq s \,.
\eeq
Let $u:\bY\to V\bY$ be a vertical vector field, with coordinate 
expression $u=u^i\der_i$, and set
\beq
\Hat{\bet} \byd i_{u_{s}}\bet \,.
\eeq
Then we have $E_{\Hat{\bet}}=e_j\vartht^j\wed\ome$,  with
\beq
e_j\ = \, J_{\up} u^i \left( \bet{^{\up}_{i}}{_j} -
\sum_{|\uq | = 0}^{k-|\up |}
(-1)^{|\up +\uq |} \frac{(\up + \uq)!}{\up ! \uq !} 
J_{\uq}\bet{^{\up + \uq}_{j}}{_i} \right) \,,
\eeq
where $0 \leq |\up | \leq s$.
\eLm

\bPf
It follows from the coordinate expression of $E_{\Hat{\bet}}$
and the Leibnitz'rule for $J_{\up}$.\QED
\ePf

\bLm
Let $\bet\in\Con{1}_{s}\wed\Con{1}_{(s,0)}\wed\Hor{n}_{s}$.
Then, there is a unique
\beq
\Tilde{H}_{\bet} \in \Con{1}_{(2s,s)}\otimes\Con{1}_{(2s,0)}\wed\Hor{n}_{2s}
\eeq
such that, for all $u:\bY\to V\bY$,
\beq
E_{\Hat{\bet}} = C^1_1 \left( u_{2s} \ten \Tilde{H_{\bet}} \right) \,,
\eeq
where $\Hat{\bet} \byd i_{u_{s}}\bet$, and $C^1_1$ stands for tensor
contraction.
\eLm

\bPf
Let $\bU\sub\bY$ be an open coordinate subset, and suppose that we have the
expression on $\bU$
\beq
\bet = \bet{^{\up}_{i}}{_j}\vartht^i_{\up}\wed\vartht^j\wed \ome \,,
\qquad
0 \leq |\up | \leq s \,.
\eeq

Then we have the coordinate expression
\beq
E_{\Hat{\bet}} = 
J_{\up}u^i
\left(
\bet{^{\up}_{i}}{_j} - \sum_{|\uq | = 0}^{s-|\up |}
(-1)^{|\up +\uq |} \frac{(\up + \uq)!}{\up ! \uq !} 
J_{\uq}\bet{^{\up + \uq}_{j}}{_i} 
\right)
\vartht^j\wed\ome \,.
\eeq

Let us set
\beq
\Tilde{H}_{\bet}[\bU] \byd
\left(
\bet{^{\up}_{i}}{_j} - \sum_{|\uq | = 0}^{s-|\up |}
(-1)^{|\up +\uq |} \frac{(\up + \uq)!}{\up ! \uq !} 
J_{\uq}\bet{^{\up + \uq}_{j}}{_i} 
\right)
\vartht^i_{\up}\ten\vartht^j\wed\ome \,.
\eeq
Then, by the arbitrariness of $u$, $\Tilde{H}_{\bet}[\bU]$ is the unique
morphism fulfilling the conditions of the statement on $\bU$.

If $\bV\sub\bY$ is another open coordinate subset and $\bU \cap \bV \neq
\emptyset$, then, by uniqueness, we have
$\Tilde{H}_{\bet}[\bU ]|_{\bU\cap\bV}=\Tilde{H}_{\bet}[\bV]|_{\bU\cap\bV}$.
Hence, we obtain the result by setting 
$\Tilde{H}_{\bet}|_{\bU} \byd \Tilde{H}_{\bet}[\bU ]$ on any coordinate open
subset $\bU \sub \bY$.\QED
\ePf

\bTh
(Generalised second variation formula). \newline
Let $\bet\in\Con{1}_{s}\wed\Con{1}_{(s,0)}\wed\Hor{n}_{s}$. Then, 
there is a unique pair of sheaf morphisms
\beq
H_{\bet} \in \Con{1}_{(2s,s)}\wed\Con{1}_{(2s,0)}\wed\Hor{n}_{2s} \,,
\quad
G_{\bet} \in \Con{2}_{(2s,s)}\wed\Hor{n}_{2s} \,,
\eeq
such that 

i. ${\pi^{2s}_{s}}^*\bet=H_{\bet} - G_{\bet}$

ii. $H_{\bet} = 1/2 \, A(\Tilde{H}_{\bet})$, where $A$ is the
antisymmetrisation map.

Moreover, $G_\bet$ is locally of the type $G_{\bet} = d_h q_{\bet}$, where 
$q_{\bet} \in \Con{2}_{2s-1}\wed\Hor{n-1}_{2s-1}$, hence $[\bet] =
[H_{\bet}]$.
\eTh

\bPf
It is clear that $G_{\bet}$ is uniquely determined by $\bet$ and
the choice $H_{\bet} = 1/2 \, A(\Tilde{H}_{\bet})$. 

Moreover, it can be easily seen \cite{Sau89} by induction on $|\up|$ that, on
a coordinate open subset $\bU\sub\bY$, we have
\beq
\bet = \bet{^{\up}_{i}}{_j}\vartht^i_{\up}\wed\vartht^j\wed \ome
= \bet{^{\up}_{i}}{_j} L_{\up}(\vartht^i)\wed\vartht^j\wed \ome
= (-1)^{|\up |}\vartht^i\wed L_{\up}(\bet{^{\up}_{i}}{_j}\vartht^j) \wed \ome
+ 2 d_h q_{\bet} \,,
\eeq
which yields the thesis by the Leibnitz' rule,
the injective morphism $\chi^s_{r}$ of remark \ref{filtration},
and the inclusions \eqref{cC and dh}
(a similar local result can be found in \cite{Bau82,Kru90}).\QED
\ePf

\bRm
In general, the section $q_{\bet}$ is not uniquely characterised.  But, if 
$\dim\bX = 1$, then there exists a unique $q_{\bet}$ fulfilling the 
conditions of the statement of the above theorem.\END
\eRm

\bCr
The presheaf $\cE_{n+1}\left(\Var{n+1}_r\right)$ is isomorphic to the image of 
the injective morphism
\beq
I_{n+2} :\left(d\For{n+1}_r/d\Thd{n+1}_r\right) \to 
\Con{1}_{4r+1}\wed\Con{1}_{(4r+1,0)}\wed\Hor{n}_{4r+1}
: [d\alp ]\mto H_{dE_{h(\alp)}}~.
\eeq
\eCr

\bPf
$I_{n+2}$ is well defined, because, recalling Lemma \ref{Helm.  equiv.}, if we 
add a suitable form $G$ of the local type $G = d_{h}q$ to $dE_{h(\alp)}$, 
the uniqueness of the decomposition of the generalised second variation 
formula (see also the above Corollary) yields $H_{G} = 0$.  
Moreover, $I_{n+2}$ is valued into
\beq
\Con{1}_{4r+1}\wed\Con{1}_{(4r+1,0)}\wed\Hor{n}_{4r+1} \sub
\Con{1}_{4r+2}\wed\Con{1}_{(4r+2,0)}\wed\Hor{n}_{4r+2}
\eeq
due to the coordinate expression of $E_{h(\alp)}$ and $H_{dE_{h(\alp)}}$; more 
precisely, being $E_{h(\alp)}$ affine with respect to the highest order 
derivatives, such derivatives disappear from the coefficient of
$dE_{h(\alp)}$  which produces the higher order coefficient of
$H_{dE_{h(\alp)}}$.  The  injectivity of $I_{n+2}$ follows from Lemma
\ref{Helm.  equiv.} and the above Corollary, because if $dE_{h(\alp)}$ and
$dE_{h(\bet)}$ fulfill 
$H_{dE_{h(\alp)}} = H_{dE_{h(\bet)}}$, then we have
\beq
dE_{h(\alp)} - dE_{h(\bet)} =  G_{dE_{h(\bet)}} - G_{dE_{h(\alp)}}\,,
\eeq
hence $[dE_{h(\alp)} - dE_{h(\bet)}] = 0$.\QED
\ePf

\bRm
Unlike the Euler--Lagrange morphism, the Helmholtz morphism is not 
characterised as being a section of a particular subsheaf.  Anyway, the 
vanishing of $[d\alp ]$ is completely equivalent to the vanishing of 
$H_{d\alp}$.  See also \cite{And86,GiMa90} for a derivation of the Helmholtz 
conditions as Euler--Lagrange equations.  Also, it is evident that the 
vanishing of $H_{d\alp}$ is a weaker condition than the vanishing of 
$d\alp$.\END
\eRm

\bCr
The sheaf morphism $\cE_{n+1}$ can be expressed via $I_{n+1}$ and $I_{n+2}$ by
\beq
\cE_{n+1}:\Var{n+1}_r\to
\Con{1}_{4r+1}\wed\Con{1}_{(4r+1,0)}\wed\Hor{n}_{4r+1} :
E \mto H_{dE}~.
\eeq
   
Moreover, if the coordinate expression of $E$ is $E = E_j \vartht^j\wed\ome$, 
then the coordinate expression of $\cE_{n+1}(E)$ is
\beq
\cE_{n+1}(E)\ = \frac 12
\left( \der{_{i}^{\up}}E_{j} - \sum_{|\uq | = 0}^{2r+1-|\up |}
(-1)^{|\up +\uq |} \, \frac{(\up + \uq)!}{\up ! \uq !} \,
J_{\uq}\der{^{\up + \uq}_{j}}E_i \right)
\vartht^i_{\up}\wed\vartht^{j}\wed\ome \,.
\eeq
\eCr

\bDf
Let $\alp \in \For{n+1}_{r}$. 

We say $H_{dE_{h(\alp)}}$ to be the {\em generalised $r$--th order Helmholtz 
morphism}.

We say $q_{dE_{h(\alp)}}$ to be a {\em generalised $r$--th order momentum\/} 
associated to the Helmholtz morphism.

We say $\cE_{n+1}$ to be the {\em generalised $r$--th 
order Helmholtz operator}.\END
\eDf

\bRm
   In this section we have obtained an intrinsic Helmholtz morphism
   that is associated to each first--order generalised Euler--Lagrange
   morphism via the sheaf morphism $\cE_{n+1}$.
   The vanishing of the Helmholtz morphism is completely equivalent to the
   standard local Helmholtz conditions (see, for example, 
   \cite{AnDu80,And86,Bau82,GiMa90,Kru90,LaTu77,Ton69}).
   
   As a by--product, to each first--order generalised Euler--Lagrange
   morphism $E\in\Var{n+1}_r$ we find a unique intrinsic contact two--form
   $G_{dE}$, where $G_{dE} = d_{h}q_{dE}$ locally; 
   $q$ plays a role analogous to that of $p$.\END
\eRm
\mysec{Inverse problems}

In this section, we show that the results of the above sections together
with the exactness of the variational sequence yield the solution for
two important inverse problems: the minimal order variationally trivial
Lagrangians and the minimal order Lagrangian corresponding to a locally
variational Euler--Lagrange morphism. As for trivial Lagrangians, our result
agrees with the local result of \cite{Gri99b,KrMu99}.

We can summarise the results of the above sections in the following theorem.
\bTh
The $r$--th order short variational sequence is isomorphic to the exact
sequence
\beq
\diagramstyle[size=2.3em]
\begin{diagram}
0 & \rTo & \R & \rTo & \For{0}_r & \rTo^{\cE_0} &
\Var{1}_r & \rTo^{\cE_1} & \dots
\\
  &      & \dots & \rTo^{\cE_{n-1}} &
      \Var{n}_r & \rTo^{\cE_n} &
      \Var{n+1}_r & \rTo^{\cE_{n+1}} &
      \cE_{n+1}\left(\Var{n+1}_r\right) & \rTo^{\cE_{n+2}} & 0~,
   \end{diagram}
\eeq
\eTh

We have two main consequences of the exactness of the above sequence.

\bCr
Let $L\in (\Var{n}_{r})_{\bY}$ such that $\cE_{n}(L) = 0$. Then, for any
$y \in \bY$ there exist an open neighbourhood $\bU\sub\bY$ of $y$ and a 
section $T\in (\Var{n-1}_{r})_{\bU}$ such that $\cE_{n-1}(T) = L$.
If $H^{n}_{\text{de Rham}}\bY =0$, then we can choose $\bU = \bY$.
\eCr

\bPf
The first statement comes from the definition of exactness for a sheaf 
sequence. The second statement comes from the abstract de Rham theorem; in 
fact, the (long) variational sequence is a (soft) resolution of the constant 
sheaf $\R$ (see \cite{Kru90,Wel80}).\QED
\ePf

\bDf
Let $L\in
(\Var{n}_{r})_{\bY}$ such that $\cE_{n}(L) = 0$. We say
$L$ to be a variationally trivial $r$--th order (generalised) Lagrangian.\END
\eDf

\bRm
If $L \in \Var{n}_r$ is variationally trivial, then 
$L$ is (locally) of the form 
$L = \cE_{n-1}(h(\alp)) = d_{h}\alp$, with $\alp\in\For{n-1}_{r}$.

We stress that a similar result is obtained in \cite{Kru93}, but 
with a computational proof.\END
\eRm

As for $(\Var{n+1}_{r})_{\bY}$, we have a result which is analogous to the 
above corollary, and justifies the following definition.

\bDf
Let $E\in (\Var{n+1}_{r})_{\bY}$. If  $\cE_{n+1}(E) = 0$, then we say
$E$ to be a locally variational (generalised) $r$--th order
Euler--Lagrange morphism.\END
\eDf

So, to any locally variational Euler--Lagrange morphism
there exists a local Lagrangian whose associated Euler--Lagrange morphism 
(locally) coincides with the given one. This is a well--known fact in the 
theory of infinite order Lagrangian sequences, but
the novelty provided by our approach is the {\em minimality\/} of the order
of the local Lagrangian. In fact, we have the following obvious proposition.

\bPr
Let $E\in (\Var{n+1}_{r})_{\bY}$ such that $E \not\in 
(\Var{n+1}_{r-1})_{\bY}$. Let $E$ be locally variational. Then, for any (local)
Lagrangian $L\in\Var{n}_{r}$ of $E$, we have $L\not\in\Var{n}_{r-1}$.
\ePr 

\bRm
In the literature there are similar results \cite{AnDu80,And86,AnTh92}, but 
proofs are done by computations. The finite order variational sequence 
provides a structural answer to the minimal order Lagrangian problem.\END
\eRm

\bRm
We stress that a minimal order Lagrangian $L \in \Var{n}_r$ for a locally
variational Euler--Lagrange morphism $E \in \Var{n+1}_r$ can be {\em
explicitly\/} computed. 

Namely, we pick an $\alp \in \For{n+2}_r$ corresponding to the
Euler--Lagrange morphism (\ie , $I_{n+1}(h(\alp)) = E$), and apply the 
contact homotopy operator (which is just the restriction of the Poincar\'e's
homotopy operator to $\Thd{n+2}_r$) to the closed form
$d\alp \in \Thd{n+2}_r$, finding
$\bet \in \Thd{n+1}_r$ such that $d\bet = d\alp$. By using once again using
the (standard) homotopy operator we find
$\gam \in \For{n}_r$ such that $d\gam = \bet - \alp$ : $L \byd I_n(\gam)$ is
the minimal order Lagrangian.

We recall that the well--known Volterra--Vainberg method for
finding a Lagrangian for $E$ yields a $(2r+1)$--th order Lagrangian.\END
\eRm
\section*{Appendix: calculus of variations}
\markboth{Appendix}{Appendix}
  \addcontentsline{toc}{section}{\hspace*{1.5em}Appendix:
  calculus of variations}

In this Appendix we give the intrinsic geometrical setting for the 
calculus of variations in Lagrangian mechanics
(\cite{GoSt73,Kru73,Tul75,Gar74,FeFr82,GaMu82,Fer83,Kol83,MaMo83b,Cos94}). The
aim is to justify the choice of the contact subsequence in the variational
bicomplex, and to give an interpretation of the results of paper.
\myskip

Suppose that a section $L\in\Hor{n}_r$ is given. Then the {\em
action\/} of an $r$--th Lagrangian $L$ on a section $s:\bI\to\bY$
($\bI$ is an orientable open subset of $\bX$  with compact closure and regular
boundary) is defined to be the real number
\beq
   \int_{\bI}(j_rs)^*L~.
\eeq

A vertical vector field $u:\bY\to V\bY$ defined on $\pi^{-1}(\bI)$ and
vanishing on $\pi^{-1}(\der\bI)$ is said to be a {\em variation field\/}.

A section $s:\bI\to\bY$ is said to be {\em critical\/} if,
for each variation field with flow $\phi_p$, we have
\beq
   \delta\int_{\bI}(J_r\phi_p\com j_rs)^*L=0~, 
\eeq
where $\delta$ is the variational derivative with respect to the parameter
$p$, and $J_r\phi_p:J_r\bY\to J_r\bY$ is first jet prolongation of the 
morphism $\phi_p$ (see \cite{MaMo83a}).

The derivative $\delta$ commutes with $\int_{\bI}$, so that the above
condition is equivalent to 
\beq
   \int_{\bI}(j_rs)^*\text{L}_{u_r}L=0
\eeq
for each variation field $u$, where $u_r:J_r\bY\to VJ_r\bY$ is the
$r$--th jet prolongation of $u$ (see the first section), and
$\text{L}_{u_r}$ stands for the Lie derivative.

Using the splitting of Proposition \ref{higher-order Lag.} (or, equivalently,
adding the form $p_{dL}$ to $L$) together with
$\text{L}_{u_r}L=i_{u_r}dL$ and the Stokes' theorem, we find that the  above
equation is equivalent to
\beq
   \int_{\bI}(j_{2r}s)^*(i_u E_{dL})=0
\eeq
for each variation field $u$.  Finally, by virtue of the fundamental lemma of 
calculus of variations, the above condition is equivalent to
\beq
   (j_{2r}s)^*E_{dL}=0~,
\eeq
or, that is the same, $E_{dL}\com j_{2r}s=0$.
\bRm 
The reason of the choice of the sheaf $\Thd{k}_r$ (for $O\leq k\leq n$) as 
the first non--trivial sheaf of the contact subsequence is now clear: for
$k=n$ $\Thd{n}_r$ is made by forms which does not contribute to the action.
   
As for the sheaf $\Thd{n+1}_r$, it is easily seen that this is precisely 
the sheaf of forms that give no contribution to the above last integral when
added to $E_{dL}$.
   
Analogously, a \lq second variation\rq\ of on Euler--Lagrange type 
operator can be defined (see \cite{Tak79}); the sheaf $\Thd{n+2}_r$ is the
sheaf of forms that give no contribution to the integral of this second
variation.\END
\eRm
\bRm
Given $\alp\in\For{n}_r$, we can extend the definition of action of a
first--order (generalised) Lagrangian $\alp$ on a section $s:\bI\to\bY$
as $\int_{\bI}(j_rs)^*\alp$. By means of a pull--back on
$J_{r+1}\bY$, we obtain the equivalent action $\int_{\bI}(j_{r+1}s)^*h(\alp )$,
being $(j_{r+1}s)^*v(\alp )=0$, and we have $h(\alp )\in\Var{n}_r$.
This explains how the $r$--th order variational sequence generalises
the $r$--th order variational calculus (see \cite{Kru90,Kru95a,Kru95b}).\END
\eRm


\end{document}